%% file: ms.tex
\documentclass[3p,twocolumn]{elsarticle} 

\graphicspath{{./figures/}}
\usepackage{graphicx,hyperref,amssymb,subcaption,amsmath,url,enumerate,placeins}
\usepackage[ruled,lined,linesnumbered,dotocloa]{algorithm2e}

\journal{Computer-Aided Design}

\biboptions{sort&compress}

\begin{document}
	
	\begin{frontmatter}
		
		\title{Object shape error modelling and simulation during early design stage by morphing Gaussian Random Fields }
		

		
		\author[1]{Manoj Babu\corref{cor1}}
		\ead{m.babu@warwick.ac.uk}
		\author[1]{Pasquale Franciosa}
		\ead{p.franciosa@warwick.ac.uk}
		\author[2]{Prashanth Shekar}
		\ead{prashant.shekhar@tufts.edu}
		\author[1]{Darek Ceglarek}
		\ead{d.j.ceglarek@warwick.ac.uk}
		
		\cortext[cor1]{Corresponding author}
		\address[1]{WMG, University of Warwick, UK}
		\address[2]{Data Intensive Studies Center, Tufts University, US}

		\begin{abstract}
			\input{abstract}

		\end{abstract}
		
		
	\end{frontmatter}
	

	\input{introduction}

\input{problemFormulation}

	\input{methodology}

	\input{caseStudy}
	\input{conclusion}
	\bibliography{extracted_2021}

\end{document}

%% file: abstract.tex
Geometric and dimensional variations in objects are caused by inevitable uncertainties in manufacturing processes and often lead to product quality issues. Failing to model the effect object shape errors, i.e., geometric and dimensional errors of parts, early during design phase inhibits the ability to predict such quality issues; consequently leading to expensive design changes after freezing of design. State-of-Art methodologies for modelling and simulating object shape error have limited defect fidelity, data versatility, and designer centricity that prevent their effective application during early design phase. Overcoming these limitations a novel \textit{Morphing Gaussian Random Field (MGRF)} methodology for object shape error modelling and simulation is presented in this paper. The MGRF methodology has (i)~high defect fidelity and is capable of simulating various part defects including local and global deformations, and technological patterns; (ii)~high data versatility and can effectively simulate non-ideal parts under the constraint of limited data availability and can utilise historical non-ideal part data of similar parts; (iii)~designer centric capabilities such as performing `what if?' analysis of practically relevant defects, and model parameters that are physically meaningful; and (iv)~capability to generate non-ideal parts conforming to statistical form tolerance specification without additional modelling effort. The aforementioned capabilities enable MGRF methodology to accurately model and simulate the effect of object shape variations on product quality during the early design phase. This is achieved by first, modelling the spatial correlation in the deviations of the part from its design nominal using Gaussian Random Field and then, utilising the modelled spatial correlations to generate non-ideal parts by conditional simulations. Practical applications of developed \textit{MGRF} methodology and its advantages are demonstrated using sport-utility-vehicle door parts.

%% file: introduction.tex
\section{Introduction} \label{sec:introduction}

Object Shape Error Modelling and Simulation (OSEMS) plays a vital role in determining the quality and functionality of mechanical products and their assembly systems. Shape error arises due to various inevitable uncertainties in manufacturing processes which lead to the actual manufactured part exhibiting imperfect form, i.e., having geometric and dimensional errors. A model of the resulting true manufactured part with geometric and dimensional errors is called a `Non-ideal part’ \citep{ISO:17450-12011}.

Object shape error, i.e., geometric and dimensional errors in non-ideal parts, leads to numerous quality related problems such as (i) high rate of re-work or scrap, (ii) inferior product functional performance, (iii) tooling failures, and (iv) unexpected production downtime which in turn, reduces both product quality and production throughput \citep{CONSAD1997}. Aforementioned critical issues necessitate accurate modelling and simulation of object shape error, especially during the early design phase. This is because accurate prediction of the effect of objects' non-ideal behaviour on the product quality during early design phase enables prevention of unnecessary design changes during later stages of production and aids achieving required product quality Right-First-Time (RFT). Thus, in-turn leading to (i) reduction of New Product Introduction (NPI) time, and (ii) elimination of the significantly higher costs of rectifying a design fault in a product or assembly system during later stages of production \citep{Haskins2004,Franciosa2014}.

The general requirements for OSEMS methodologies have been discussed in existing literature  \citep{Yan2018}. However, a detailed analysis of the requirements from an early design perspective has not been performed yet. Therefore, we first describe the essential requirements for OSEMS methodologies during early design; they can be broadly classified into the following categories: 

\begin{enumerate}
	\item \textit{Object fidelity} - refers to the ability of the methodology to be applied to objects/parts with varying levels of surface complexity. These surfaces in increasing order of complexity can be classified into the following three types, (i) planar or 2D surfaces; (ii) 3D primitives such as spherical, cylindrical, and conical surfaces; and (iii) complex 3D free form surfaces. 
	
	While planar and 3D primitive surfaces are fundamental blocks essential for the functioning of many mechanical assemblies, complex 3D free form surfaces due to their high functional and aesthetical utilities are increasingly employed in automotive, aerospace and optics industry \citep{Savio2007a}. Therefore, a methodology should be able to model shape error of all the three types of surfaces.
	
	\item \textit{Defect fidelity} - refers to the ability of the methodology to accurately emulate various types of part defects encountered during the manufacturing process. This requirement is vital to accurately model the assembly process and enable the diagnosis of various complex and  ill-conditioned faults arising during the assembly process of compliant parts and to enable to achieve the required assembly quality RFT.
	
	High defect fidelity can be achieved by (i) simulating shape error at multiple spatial scales, i.e., local, and global deviations  \citep{Babu2018,Luo2018}; and (ii) simulating technological patterns, i.e., spatial deviation patterns specific to a given manufacturing process. 
	
	Global deviations are the non-ideal behaviour affecting a large area or the entire part. In contrast, local deviations are those that affect a small or localised region of the part. In sheet metal parts fabricated by forming, shape error caused by spring back could affect the entire part and can be classified as a global deviation. Whereas surface dents or flange deformations that typically affect a localised region can be classified as local deviations. The simulation of global and local deviations is important because some manufacturing processes require tighter dimensional quality requirement in a localised region or have variable tolerance requirements throughout the part. For instance, in a sheet metal assembly, the non-ideal behaviour of part flanges is critical as fastening/joining of parts takes place along the flange. Any variation in this region has a high probability of affecting the key product characteristics and thus needs to be simulated and analysed during design.
	
	On the other hand, simulation of technological patterns enables the creation of a non-ideal part that captures the spatial deviation patterns specific to a given manufacturing process. The non-ideal part thus simulated is a better representation of a true manufactured part and hence can provide accurate assembly simulation and tolerance analysis results compared to non-ideal part with random form variations. Therefore, a methodology should possess high defect fidelity while simultaneously satisfying other criteria in described in this section.
	
	\item \textit{Data versatility} - refers to the ability of the methodology to perform effectively at different levels of data availability and to handle the imprecision and uncertainty arising during various stages of NPI. Imprecision relates to the constantly evolving object geometry, especially during early stages of NPI. Whereas uncertainty relates to the various levels of part-to-part variation occurring during different stages, such as from initial prototyping to final product and full production.
	
	For instance, during the NPI process of an assembly system development the various stages are engineering, manufacture and assembly, installing and commissioning, launch, and production. During this NPI journey from engineering to production stage, availability of measurement data or data from manufacturing process simulations from which the non-ideal behaviour of parts can be quantified varies from, (i) no data, (ii) historical data from similar/surrogate parts, (iii) preproduction data from same product, to (iv) data on production parts. 
	
 	A methodology to be data versatile should be able to effectively handle this imprecision and uncertainty, and generate non-ideal parts at any of the above-described levels of data availability. Especially during the early design phase where limited data is available (stages (i) through (iii) described in the previous paragraph) in order to prevent unnecessary design changes during later stages of production and eliminate the corresponding time and cost penalties associated with it.
	
	\item \textit{Designer centricity} - refers to the methodology’s ease of applicability, and the interpretability of its parameters by the designer. During early stages of design, ease of applicability can be understood as ability to simulate technologically wanted design variations to perform several `what if?’ analysis with little or no additional modelling effort. Examples of `what if?’ analysis could include the ability to quantify the effect of the following non-ideal behaviours on assembly quality: (i) various levels of spring-back, (ii) specific local part deformations such as flange variations and dents, and (iii) bending or twisting about a given axis. The ability to perform these `what if?’ analysis early during design with little or no additional modelling effort enables timely fault detection and helps prevent expensive design changes during later stages.

	On the other hand, interpretability of model parameters criterion enables the designer to understand/evaluate the model parameters with ease and to manually tune them - especially when no historical or simulation data are available to aid optimum parameter identification. The interpretability of model parameters can be improved if (i) their dimensions are in the modelling space as opposed to low dimensional or latent space, and (ii) they are physically meaningful. A methodology should possess the  aforementioned qualities to be considered designer centric.

	\item \textit{Support for tolerance analysis and synthesis}- refers to the ability to generate non-ideal parts conforming to statistical form tolerance specification without additional modelling effort. In this study we focus specifically on form tolerance for profile of a surface  \cite{ISO-TC2013}, as most other types of tolerance specifications can be simulated through rigid transformations or scaling of the resulting part. The ability to generate non-ideal parts that resemble real manufactured parts and conform to tolerance specifications enables to perform accurate assembly process simulations and thus helps to allocate optimum tolerances early during the design.

	\item \textit{Computational intensity and scalability} – refer to the ability of the methodology to be (i) computationally less intensive, and (ii) scalable to assembly with large number of parts. While the necessity of being computationally less intensive is easily understood, the necessity to be scalable to large assemblies stems from the fact that most mechanical assemblies have a large number of parts - for instance an automotive body assembly process is a multi-levelled hierarchical process in which 200-250 sheet metal parts are assembled together to form the final product \citep{Shiu1996, Shi2006a} - and the methodology should be capable of being applied to them. 
	A key factor that can affect scalability is the total number of model parameters. This is because assembly system optimisation by taking into account compliant parts’ non-ideal behaviour is typically performed using computationally expensive Finite Element Method (FEM) simulations \citep{Liu1997,Shahi2018} and the number of FEM simulations required increases exponentially with the number of model parameters whose effect on product quality has to be determined. Thus, a design friendly model is the one with a small number of model parameters which reduces the number of computationally expensive FEM simulations necessary.

\end{enumerate}

Though numerous methodologies to model object shape error exist \cite{Huang2004,Sarraga2004,Samper2007,Sarraga2010,Franciosa2010, Zhang2012,Sanchez-Reyes2012,Zhang2013, Huang2014,Schleich2014c,Das2016a,Homri2017, Zhang2017,Luo2018,Liu2018,Luo2020}, they exhibit several limitations with respect to the aforedescribed criteria such as (i) limited defect fidelity due to which they are unable to accurately simulate many true non-ideal part behaviours, (ii) limited data versatility due to which they are unable to operate effectively at different levels of data availability often found during early design phase, and (iii) limited designer centric capabilities due to which they are unintuitive for designers.

Overcoming these limitations, this paper presents the Morphing Gaussian Random Field (MGRF) methodology to model and simulate object shape error, primarily during early design phase. The contributions of this study are the development of (i) a high defect fidelity OSEMS methodology capable of simulating local and global deformations, and technological patterns; (ii) an OSEMS methodology that has high data versatility and can effectively simulate non-ideal parts at all levels of data availability; (iii)  a highly designer centric OSEMS methodology capable of performing `what if?' analysis, and with model parameters that are physically meaningful; and (iv) an OSEMS methodology capable of generating non-ideal parts conforming to statistical form tolerance specification without additional modelling effort.

The rest of the paper is organised as follows: In Section \ref{sec:litReview}, we classify the existing OSEMS methodologies, analyse their applicability with respect to the criteria described in Section \ref{sec:introduction}. In Sections~\ref{sec:problemFormulation} and \ref{sec:method} we describe the problem formulation and the developed MGRF methodology in detail, respectively. In Section \ref{sec:Case1}, we demonstrate the MGRF in various scenarios using automotive door inner parts. Finally, in Section \ref{sec:conclusion} we discuss the conclusions and future research.

\section{Literature review}\label{sec:litReview}

State-of-art OSEMS methodologies can be classified into two main categories \citep{Yan2018} (i) morphing based, and (ii) deviation decomposition or mode-based. Morphing based methodologies model non-ideal parts by modifying either a parametric geometry represented by Bezier, NURBS, and B-spline; or discrete geometry represented by mesh or Cloud-of-Points (CoP). Whereas deviation decomposition or mode-based methodologies model non-ideal parts by decomposing part deviations from measurement or simulation data into a linear combination of orthogonal modes, and are typically applied to discrete geometry representations.

A key concept in addition to the two categories of classification described above is that of the Skin model. It is defined as a model of the physical interface between the workpiece and its environment \citep{ISO:17450-12011} and is based on the tenants of GeoSpelling, a coherent uni-vocal language for non-ideal part specification and verification \citep{Ballu2015}. All existing OSEMS methodologies can be considered as different means to generate skin model shapes, i.e., unique finite skin model representatives comprising of deviations from ideal manufacturing and assembly process \citep{Schleich2014c}.


\subsection{Morphing based OSEMS methodologies} \label{ssec:morph}

Volume splines have been used to fit deformed or deviational point data to CAD and find the non-ideal part by minimizing a sum of squared error function in  \citep{Sarraga2004,Sarraga2010}. 
Hermite approximation was applied to reduce the high degree polynomial compositions when an explicit expression of the surface is needed during free-form deformation in \citep{Sanchez-Reyes2012}. A NURBS based interpolation of measurement data focused on the reconstruction of CAD geometry with form errors was proposed in \citep{Zhang2017}. While the afore-discussed techniques mainly focus on reconstruction of CAD geometry for accurate part representation, the Envelope-T model to simulate non-ideal parts for tolerance analysis was proposed in \citep{Luo2018,Luo2020}. 

A morphing mesh methodology based on constrained deformation was employed to generate non-ideal part in \citep{Franciosa2010}. 
Second order shapes have been used to model the systematic deviations of discrete geometry representation of non-ideal parts in \citep{Zhang2012}.

\subsection{Deviation decomposition or mode-based OSEMS methodologies} \label{ssec:devDecomp}

Statistical Modal Analysis (SMA) a mode-based methodology was proposed in \citep{Huang2014}, where, shape deviation is decomposed into a set of orthogonal patterns  based on Discrete Cosine Transform (DCT). Geometric Modal Analysis (GMA) a technique utilising 3D-DCT, capable of characterising shape variations in 3D surfaces was proposed in \citep{Das2016a}.

Natural Mode Analysis, based on the decomposition of deviation data into natural modes of vibration was developed in \citep{Samper2007}. Metric Modal Decomposition (MMD) methodology based on  a variation of Natural Mode Decomposition, was proposed in \citep{Homri2017}. Principal Component Analysis (PCA) is used to characterise part variation in \citep{Zhang2013}. A random field-based methodology to model and simulate part shape error was developed in \cite{Schleich2014c}.


\subsection{Applicability of state-of-art OSEMS methodologies during early design stage}\label{ssec:applicability}

The State-of-Art OSEMS methodologies have many advantages such as designer centricity and computational advantages of morphing mesh methodology due to its physically meaningful model parameters low computational requirements \cite{Franciosa2010}. The ability to work under the constraint of limited or no data availability of a few morphing based methodologies \cite{Franciosa2010,Sanchez-Reyes2012,Zhang2017,Luo2018,Luo2020}. Support for statistical form tolerance without additional modelling effort by Envelop-T methodology \cite{Luo2018,Luo2020}. Technological patterns modelling capability of mode-based methodologies due to their ability to learn from measurement and simulation data \cite{Huang2014,Das2016a,Samper2007, Zhang2013,Schleich2014c,Homri2017}.

However, they also entail many limitations such as the morphing mesh methodology's  \cite{Franciosa2010} inability to model technological patterns. Limited defect fidelity and designer centricity of morphing based methods \cite{Sanchez-Reyes2012,Zhang2017,Luo2018,Luo2020} as they  (i)~require setup and optimisation of many parameters such as degree of polynomial, location of knots, numbers of spline intervals and displacement vectors; (ii)~lack the ability to model spatial deviation patterns; and (iii)~face non-trivial challenges to demarcate the region of control influencing the local deformation, as it depends on cell dimensions and spline degree which are difficult to predict and modify because it requires an expert knowledge of the underlying mathematics making them unfriendly to designers \cite{Gain2008}. Limited defect fidelity, data versatility and designer centricity of mode-based methodologies \cite{Huang2014,Das2016a,Samper2007, Zhang2013,Schleich2014c,Homri2017} due to their inability to model local deformations, model parameters typically being present in frequency space or other reduced dimension latent space, high computational requirements, and inability to perform effectively when measurement and simulation data are not available \cite{Yan2018}. A summary of the aforediscussed  analysis of the applicability of State-of-Art OSEMS methodologies to early design phase according to the criteria detailed in Section \ref{sec:introduction} is presented in Table~\ref{nonIdealMtdClas}. 

Therefore, overcoming these limitations we develop the MGRF methodology that satisfies all the criteria described in Section \ref{sec:introduction}. A detailed description of the problem formulation and the developed MGRF methodology are presented in Sections~\ref{sec:problemFormulation} and~\ref{sec:method} below.

\begin{table*}[htb]
	\caption{Summary of literature review of state-of-art non-ideal part shape error modelling methodologies}
	\label{nonIdealMtdClas}
	\centering
	\includegraphics[width=\textwidth]{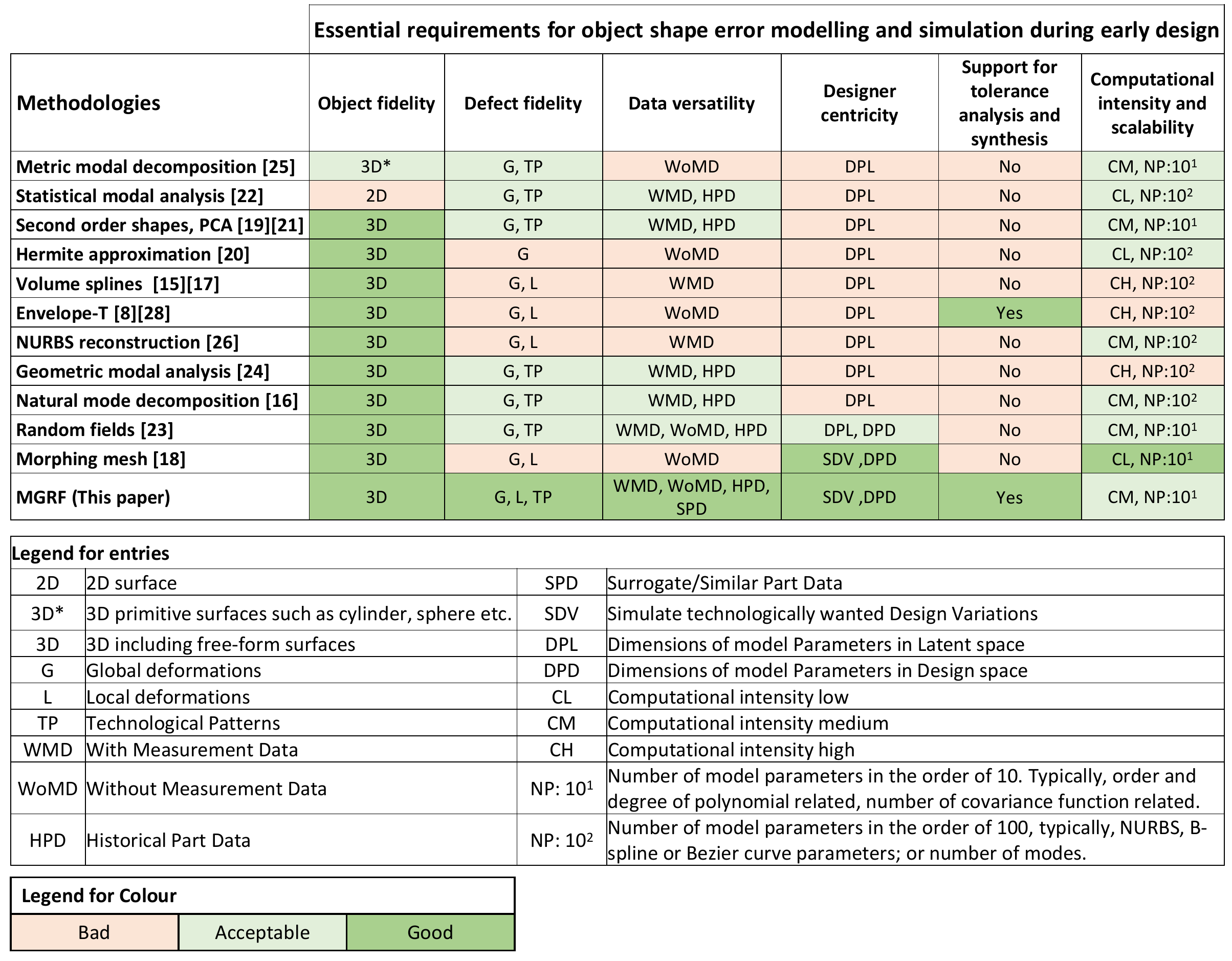} 
\end{table*}

%% file: problemFormulation.tex
\section{Problem formulation} \label{sec:problemFormulation}

Representing the nominal coordinates of points on the surface of the part by $ \mathbf{X} $, a $\textrm{N}\times3 $ matrix, where, N is the number of nodes in the mesh representation of the part; the non-ideal deviation of a part along the surface normal direction, represented by $Z$, is modelled as a Gaussian Random Field (GRF), $ f $. A GRF is a collection of random variables $\{Z_\mathbf{x}\}$ indexed by a set $\{\mathbf{x}|\mathbf{x}\in \mathbb{R}^n\}$, where any finite subset of $ Z $'s has a joint multivariate Gaussian distribution \cite{Rasmussen2006gaussian,ChilesJeanBook,Adler2010,Lantuejoul2013geostatistical}. The GRF provides a means to generate spatially correlated random variables which are well suited to model non-ideal part behaviour, and overcomes the limitations of 1D-Gaussian and multi-Gaussian methods that could generate unrealistic shapes \cite{Schleich2014c}.

The GRF forms a functional relationship between the nominal location $ \mathbf{X} $ and the non-ideal deviation $ Z $, i.e, $ f:\mathbf{X} \rightarrow Z $; and is completely characterised by the mean function and the covariance function which affect the shape and smoothness of the generated non-ideal part \cite{Adler2010}. The mean function is defined as $ \mathbb{E}[f( \mathbf{x})] $,  where, $ \mathbb{E} $ is the expectation operator, can be estimated by fitting regression models \citep{cressie2015statistics}. Since an ideal part's deviation from nominal is zero, in the present study mean function is taken to be zero without affecting the analysis \citep{Rasmussen2006gaussian}. 



The covariance between two nodes $  \mathbf{x}, \mathbf{x'}$ on the mesh representation of the part can be represented by the covariance function, $ C( \mathbf{x}, \mathbf{x'}) $. For a function to represent the covariance between two input points, it has to be positive semi-definite \citep{Vanmarcke1983}. Various types of covariance functions with different characteristics exist, a detailed review is presented in \citep{Rasmussen2006gaussian}. 

\begin{figure*}[htb] 
	\centering	
	\includegraphics[width=\textwidth]{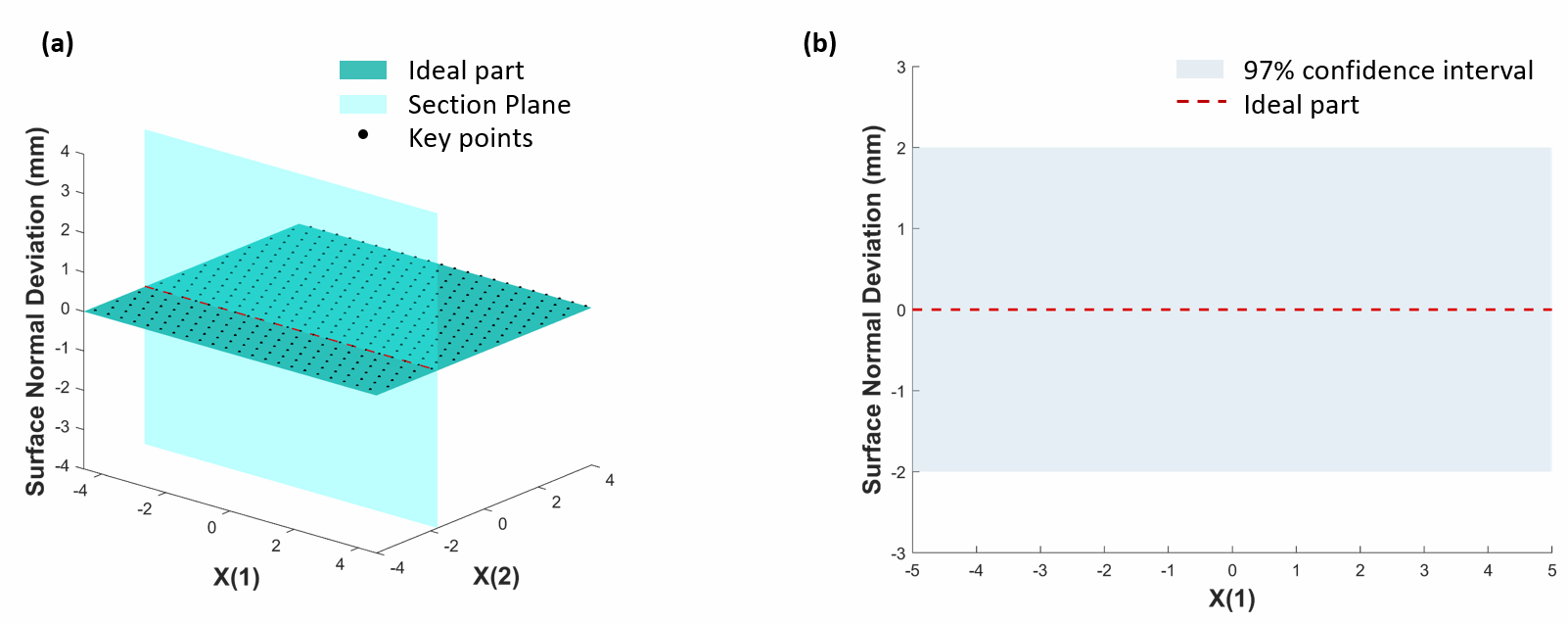}
	\caption[Illustration of a 2D ideal part with key points]{ {Illustration of a 2D ideal part with key points; (a) 2D Illustration with key points, (b) Cross-sectional view}}
	\label{fig:ideal1Dn2D}
\end{figure*}

The spatial pattern of the simulated non-ideal part and the smoothness of deviations from nominal depend on the parameters of the covariance function. To illustrate the effect of change in covariance function parameters on the simulated non-ideal part spatial patterns we utilise the squared exponential covariance function represented by Eq.~\eqref{eq:squaredExp},
\begin{equation}
\label{eq:squaredExp}
C_{SE}(\mathbf{x},\mathbf{x'})=\sigma^{2}_{f}\exp\left[-\frac{1}{2}\sum_{i=1}^{D}\frac{(x_i-x'_i)^2}{l_i^2} \right]
\end{equation}
where, $\sigma_f^2$  is the scaling factor, $ D$ is the dimension of the input space, i.e.,~2 for the illustration in Fig.~\ref{fig:ideal1Dn2D} and $3$~for the case of 3D-non-ideal part modelling, $ l_i$ is the correlation length along dimension $ i$. The scaling factor is equivalent to the variance of the GRF and can be utilised to simulate non-ideal parts conforming to a given statistical tolerance specification as described in Section~\ref{ssec:predVariation}. 

\begin{figure*}[htb]
	\includegraphics[width=\textwidth]{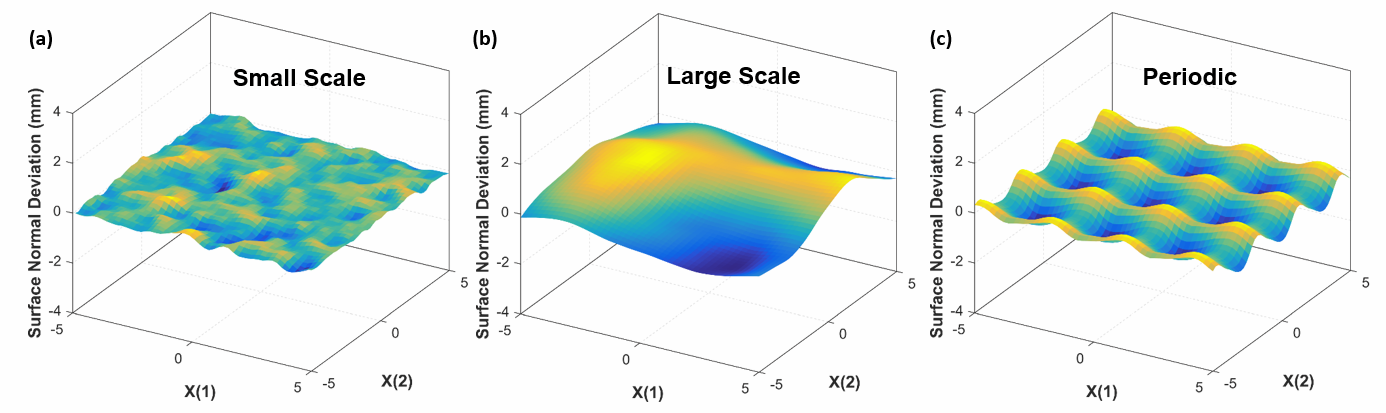} 
	\caption{Illustration of multi-scale and pattern modelling abilities}
	\label{fig:covScale}
\end{figure*}

When the correlation length is small, the deviations of two points which are at a distance longer than the correlation length are independent of each other, and the resulting non-ideal deviation pattern can simulate roughness, as illustrated in Fig.~\ref{fig:covScale}a. Similarly, when the correlation length is large, it simulates smoother form variation as shown in Fig.~\ref{fig:covScale}b. Periodic errors can be simulated using Eq.~\eqref{eq:perCov},
\begin{equation}
\label{eq:perCov}
C_{per}(\mathbf{x},\mathbf{x'})=\sigma^{2}_{f}\exp\left[-\frac{1}{2}\sum_{i=1}^{D}\left(\frac { \sin \left(\frac{\pi (x_i-x'_i)}{p_i}\right)}{l_i} \right)^2 \right]
\end{equation}
a modified squared exponential covariance function, where, $ p_i$ represents the periodicity of the pattern along dimension $ i $ \citep{Mackay1998introduction}. A simulated periodic pattern from Eq.~\eqref{eq:perCov} is illustrated in Fig.~\ref{fig:covScale}c. Therefore, varying the parameters of the covariance function enables the simulation of multi-scale errors, such as roughness, waviness, and large scale form errors. The correlation lengths influence the extent to which a given deviation affects the neighbouring points in a given direction, and are conceptually equivalent to the domain of influence in the morphing mesh methodology \citep{Franciosa2010} which defines the region of influence of the deviation.

\begin{figure}[htb] 
	\centering	
	\includegraphics[width=0.5\textwidth]{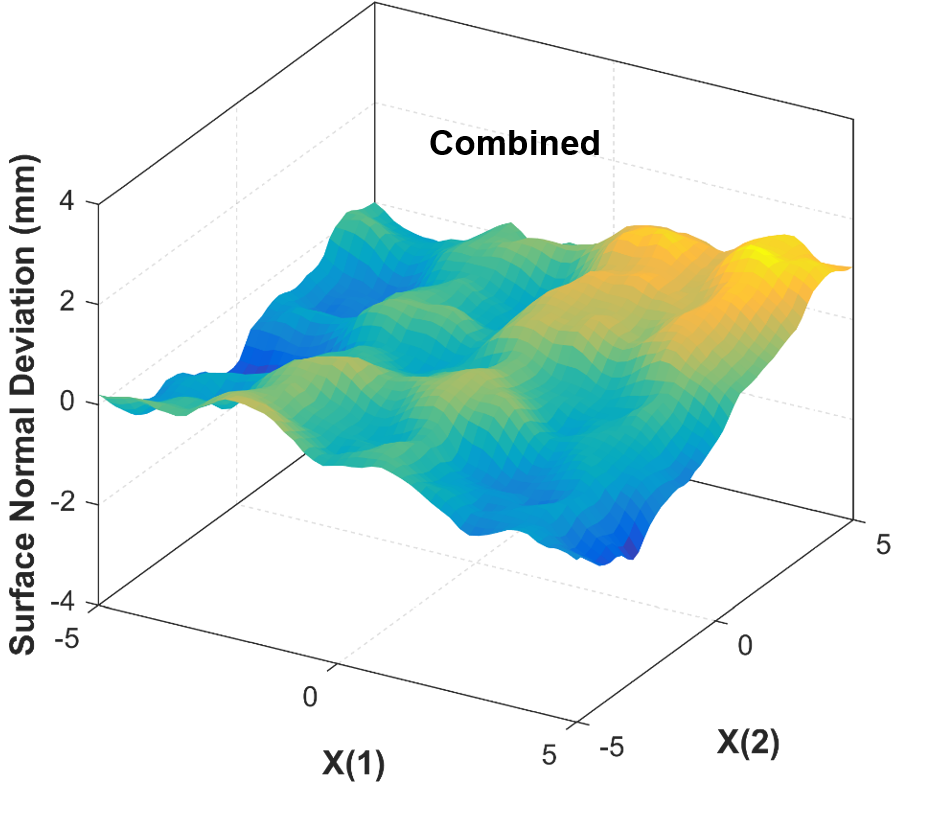}
	\caption{An illustration of a non-ideal part with complex pattern}
	\label{fig:sum}
\end{figure}

Various covariance functions, such as periodic and squared exponential or two squared exponential covariance functions with different length and scale parameters can be combined to form a valid covariance function as in Eq.~\eqref{eq:covSum} \citep{Rasmussen2006gaussian}.
\begin{equation}
\label{eq:covSum}
C_{sum}(\mathbf{x},\mathbf{x'})=C_{SE_1}(\mathbf{x},\mathbf{x'})+C_{SE_2}(\mathbf{x},\mathbf{x'})+C_{per}(\mathbf{x},\mathbf{x'})
\end{equation}
This property of covariance functions enables simulation of complex patterns in non-ideal parts, an illustration of a non-ideal pattern simulated from a combination of small scale $ (C_{SE_1}) $, large scale $ (C_{SE_2}) $ and periodic $ (C_{per}) $ covariance function is shown in Fig.~\ref{fig:sum}.

A non-ideal part with a combination of different deviation patterns can be decomposed into its constituents by using the combined covariance function, and the optimum parameters can be automatically estimated by maximising the marginal likelihood as described in Section~\ref{ssec:optimiseParam}. The aforementioned ability of the covariance function to automatically model and extract complex deviation patterns \citep{Duvenaud2013}, along with; (i)~the asymptotic properties of mean of covariance function parameters obtained in many settings through maximum likelihood estimation being equal to the true values of the parameters; and, (ii)~the empirical variogram's  inability to exactly define the differentiability of the random field  \citep{Stein2012interpolation}, make the modelling of non-ideal part deviations using covariance function advantageous compared to the variogram traditionally used in geostatistics.

In this section, to simplify exposition of the simulation of non-ideal parts we utilise a 2-Dimensional (2D) part whose ideal form is the flat plate illustrated in Fig.~\ref{fig:ideal1Dn2D}a. A cross-sectional view of the flat plane is illustrated in Fig.~\ref{fig:ideal1Dn2D}b. 

\begin{figure*}[htb] 
	\centering	
	\includegraphics[width=\textwidth]{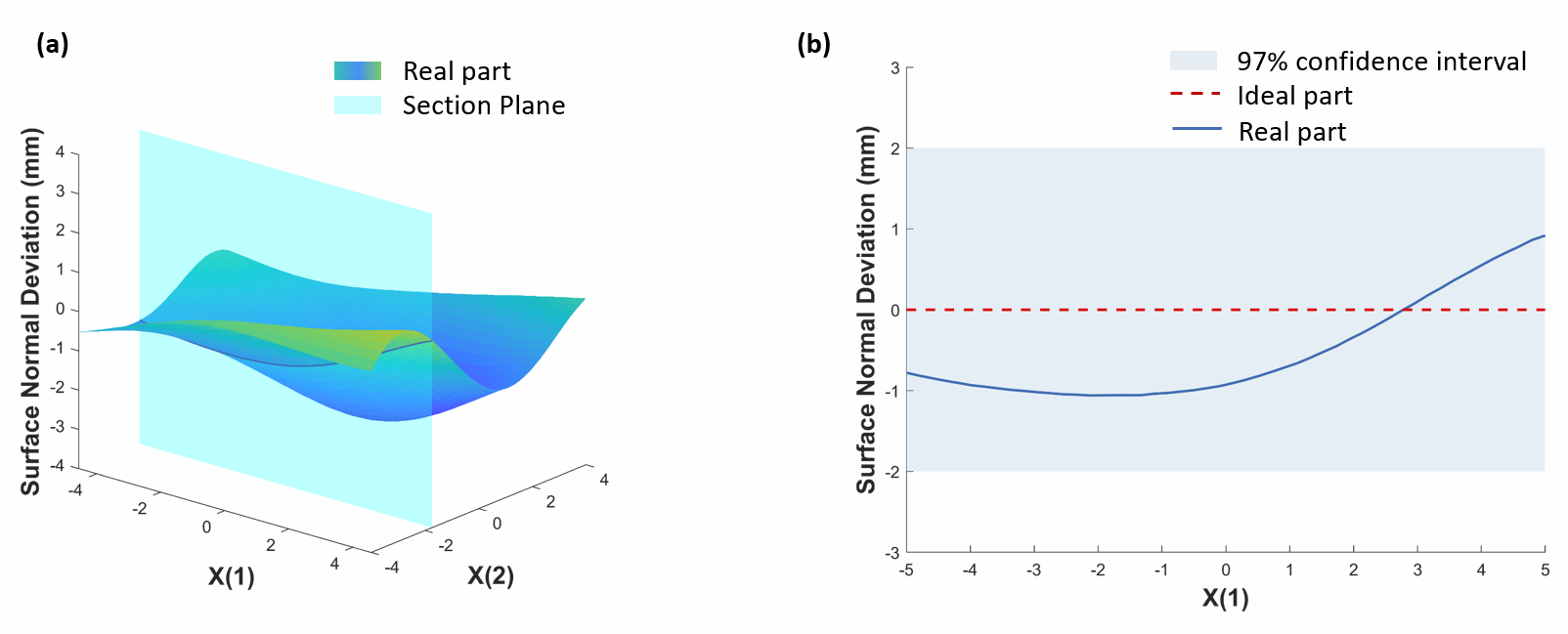}
	\caption[Illustration of a 2D real part]{ {Illustration of a 2D real part; (a) 2D Illustration, (b) Cross-sectional view}}
	\label{fig:real1Dn2D}
\end{figure*}

An illustration of a real manufactured flat plate with form error (exaggerated for illustration purposes) is shown in Fig.~\ref{fig:real1Dn2D}. The spatial deviation pattern of a given part from its design nominal is characterised by estimating the optimum values of the parameters of the covariance function $ \theta = \{\sigma_f^2, l_i,p_i \} $, that best describe it. While the methodology to obtain the optimum parameters' value and generate non-ideal parts that exhibit similar spatial deviation patterns as the real manufactured part are described in detail in Sections~\ref{ssec:modelling}~and~\ref{ssec:simulation}, in this section we illustrate the key concepts of unconditional simulation and morphing GRF through conditional simulation using the flat plate.

Unconditional simulation is a spatially consistent Monte-Carlo simulation with the goal to mimic spatial variation pattern as realistically as possible. Whereas, conditional simulation in addition to mimicking spatial variation pattern constrains the simulated non-ideal parts to pass through given data points or fixed key points \citep{ChilesJeanBook,Lantuejoul2013geostatistical}. 

\begin{figure*}[htb] 
	\centering	
	\includegraphics[width=\textwidth]{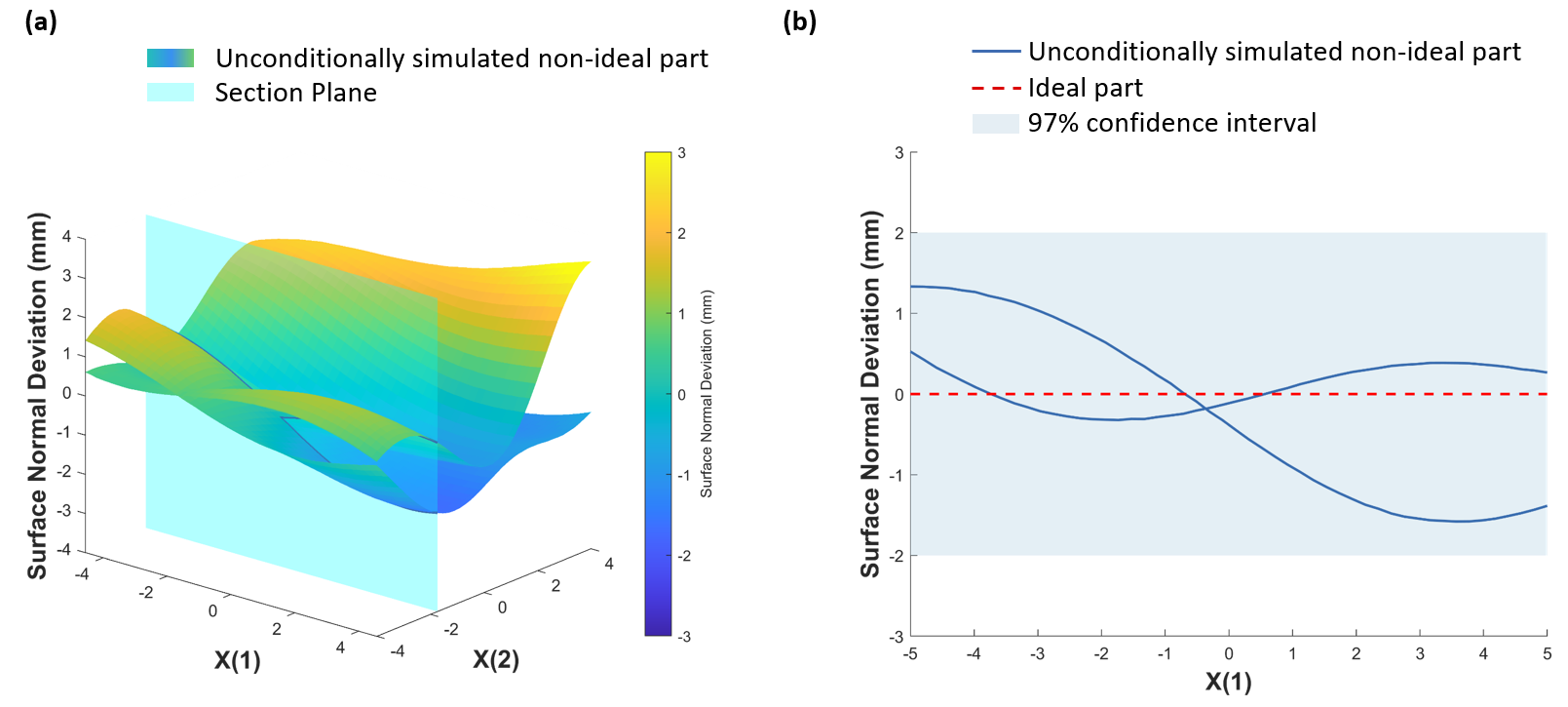}
	\caption[Illustration of unconditional non-ideal part simulation]{ {Illustration of unconditional non-ideal part simulation; (a) Two simulated parts, (b) Cross-sectional view}}
	\label{fig:unCond1D}
\end{figure*}

Though the non-ideal parts generated by unconditional simulation emulate the spatial deviation patterns of manufactured part it cannot incorporate design intent or convey any physical meaning. Additionally, the generated non-ideal parts are equally likely to take any shape around the mean. An illustration of two non-ideal parts generated by unconditional simulation for a statistical tolerance limit of $ \pm2 $ mm  with 97\% confidence interval (CI) is shown in Fig.~\ref{fig:unCond1D}. It shows that the simulated non-ideal parts emulate deviation pattern of the source, i.e., long correlation length causing smooth deviations. 

\begin{figure*}[htb] 
	\centering	
	\includegraphics[width=\textwidth]{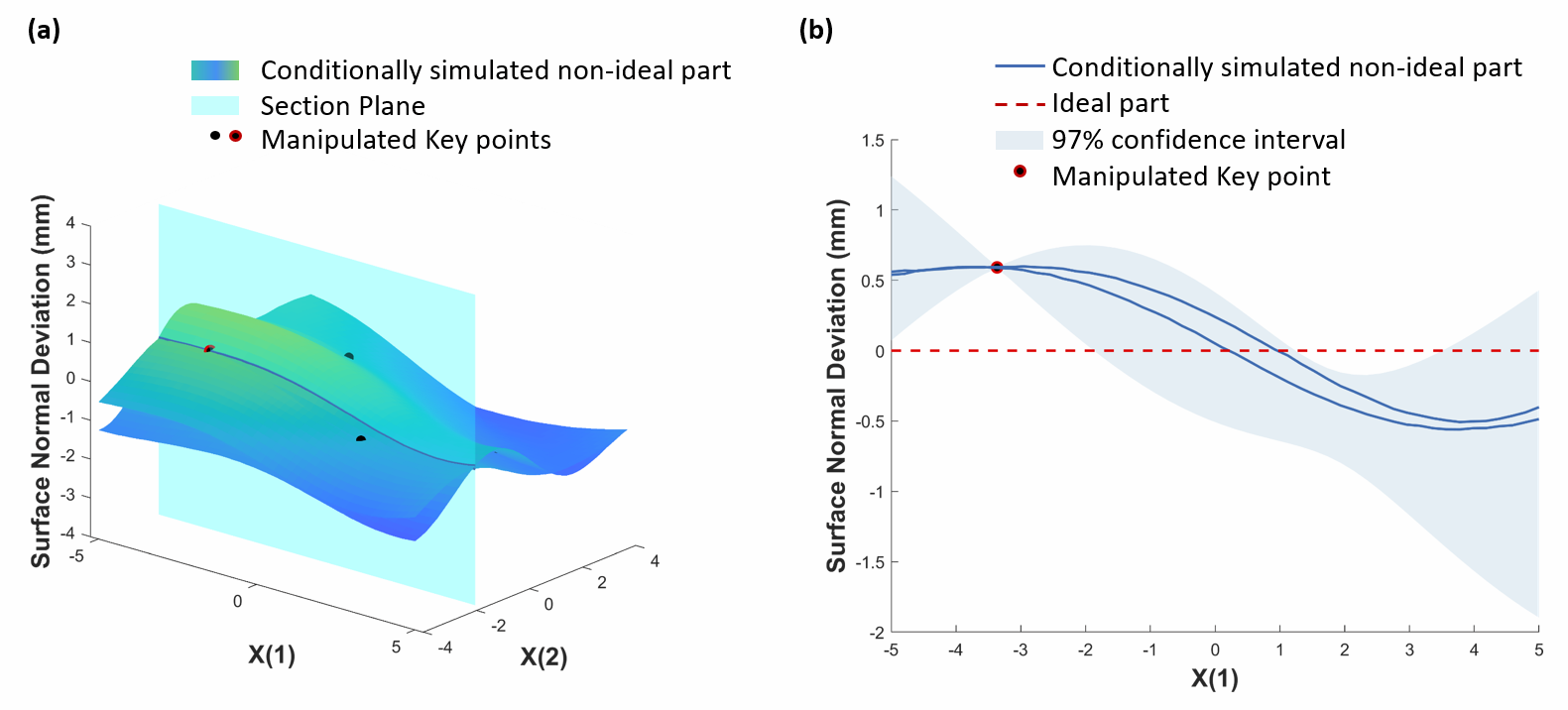}
	\caption[Illustration of 2D non-ideal parts generated by MGRF methodology]{ {Illustration of 2D non-ideal parts generated by MGRF methodology; (a) Two simulated parts, (b) Cross-sectional view}}
	\label{fig:cond1D}
\end{figure*}

The ability to incorporate user knowledge and simulate physically meaningful global and local non-ideal part deviations is achieved by MGRF. Morphing is performed by manipulating a subset of key points on the surface of the part and generating non-ideal parts that pass through them by conditional simulation. Key points are specific points on the surface of the part which the designer can manipulate to simulate parts conforming to given design intent. Key points are chosen to uniformly cover the surface of the part as illustrated in Fig.~\ref{fig:ideal1Dn2D}a, a more detailed description of key points is presented in Section~\ref{ssec:keyPointSel}. An illustration of two non-ideal parts simulated through MGRF by manipulating three key points is shown in Fig.~\ref{fig:cond1D}a, the simulated parts pass through the manipulated key points. This process of simulating non-ideal parts that conform to design intent by constraining them to pass thorough a few manipulated key points is called MGRF. The magnitude of deviation of key points from nominal is set by the designer in accordance with the design requirements as detailed in Sections~\ref{ssec:predVariation} and~\ref{ssec:dataCase}. As illustrated in Fig.~\ref{fig:cond1D}b, the tolerance bounds for constrained non-ideal parts are smaller than that of unconditional non-ideal parts because the shape variation of a generated non-ideal part near the fixed key point is limited due to geometric covariance \citep{Merkley1998}. This constraint eases as we move away from the key point, accordingly, the uncertainty in the shape of the predicted non-ideal part is low near the key point and increases as we move further away from it. A detailed explanation of the methodology to generate non-ideal parts by MGRF is provided in Section~\ref{sec:method} below.

%% file: methodology.tex
\section{Methodology} \label{sec:method}

The MGRF methodology proposed to model and simulate shape error of parts has two main stages (i)~non-ideal part modelling, and (2)~non-ideal part simulation. Firstly, the parameters of the GRF characterising the deviation of given non-ideal parts are estimated during the non-ideal part modelling stage. Following the non-ideal part modelling stage, parts that emulate the spatial deviation pattern of the manufactured part and conform to design intent are generated by MGRF through conditional simulation in the non-ideal part simulation stage. The two main stages are explained in detail in Sections~\ref{ssec:modelling}~and~\ref{ssec:simulation} below.

\subsection{Non-ideal part modelling} \label{ssec:modelling}
\begin{figure}[htb]
	\centering
	\includegraphics[width=0.35\textwidth]{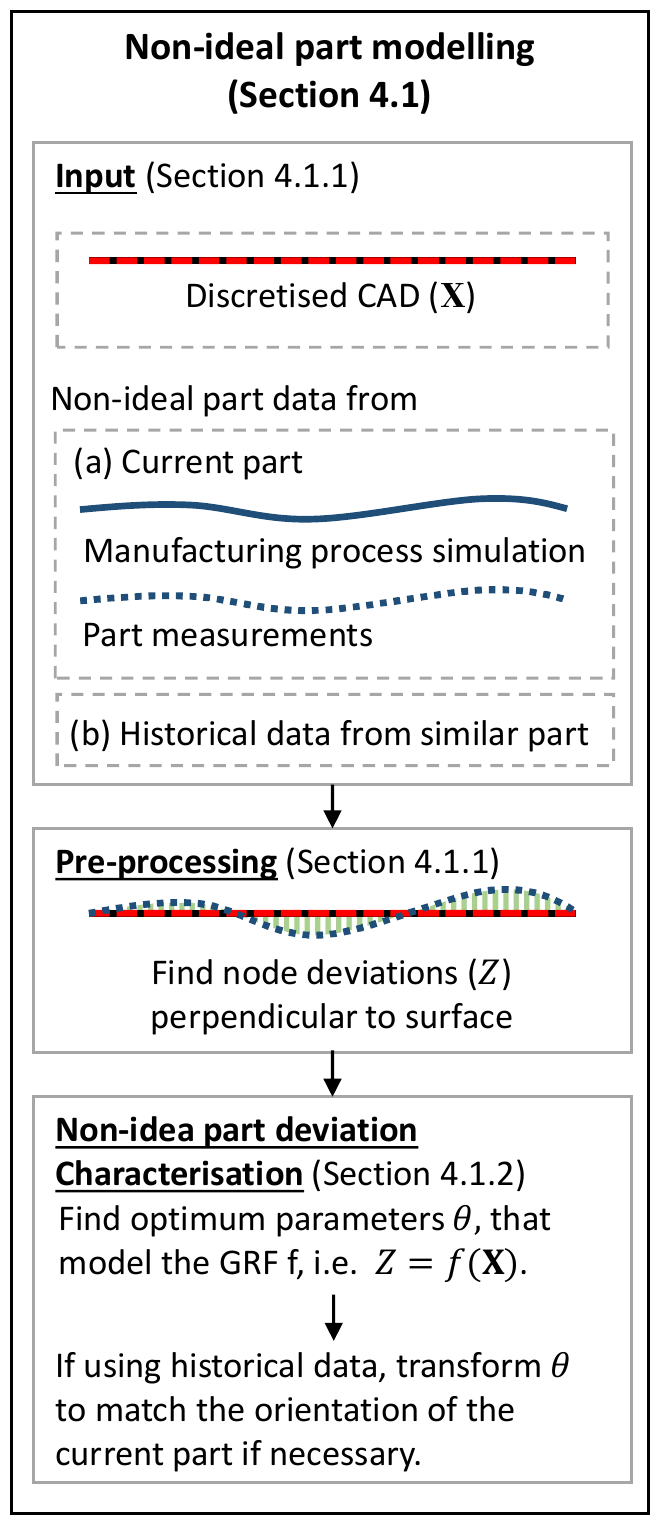} 
	\caption{Schematic representation of non-ideal part modelling methodology}
	\label{fig:nonIdealModel}
\end{figure}

Non-ideal part modelling stage estimates the covariance function parameters of the GRF characterising the spatial deviation pattern of a given manufactured part. Representing by $ \theta = \{\sigma_f^2, l_i^{-2}, p_i\}$, the set of all covariance function parameters, the aim of non-ideal part deviation modelling is to find the optimum $ \theta $, that best describe the given non-ideal part deviation. Non-ideal part modelling consists of (i)~input Pre-processing, and (ii)~non-ideal part deviation characterisation. The non-ideal part modelling stage is schematically illustrated in Fig.~\ref{fig:nonIdealModel} and is explained in detail in Sections~\ref{ssec:input}~and~\ref{ssec:optimiseParam}.

\subsubsection{Input and pre-processing} \label{ssec:input}

The inputs for the proposed methodology are (i)~the Computer Aided Design (CAD) geometry or mesh file of the ideal part, and non-ideal part data from measurement or simulation of current part, or (ii)~historical data of similar part consisting of mesh file and non-ideal part data. 

Pre-processing involves, calculating the surface normal deviation of the mesh nodes from available non-ideal part data. The mesh node coordinates of the nominal part represent the set of all points ($ \mathbf{X} $) on the surface of the part. If measurement data (typically Cloud-of-Points (CoP)) are available, $Z$, the deviation of each node from the design nominal in the surface normal direction is calculated by finding the distance between each ideal mesh node and the aligned CoP using routines provided in \cite{Franciosa2018}. If data of non-ideal part deviations are available through manufacturing or assembly process simulation, for instance from FEM analysis, the node deviations from simulation can be projected along the surface normal direction to obtain $Z$. Following the estimation of $ Z $, optimum covariance function parameters $ \theta $ are obtained as described in the following section.

\subsubsection{Non-ideal part deviation characterisation} \label{ssec:optimiseParam}

The ability to mimic the spatial deviation pattern of the true manufactured part as closely as possible depends on the estimation of covariance function parameters $ \theta $, that best describe a given non-ideal part deviation. Optimum values of $ \theta $ are found by maximising the log-likelihood of covariance function parameters given the part surface normal deviation data $ {Z} $, estimated in Section~\ref{ssec:input}. The log-likelihood function for a GRF is a function of $\theta$ for a given non-ideal deviation $Z$, and is estimated using Eq.~\eqref{eq:margLik} \citep{Stein2012interpolation,Rasmussen2006gaussian}. 
\begin{equation} 
\ln(L(\theta|{Z}))=-\frac{N}{2}\ln(2\pi)-\frac{1}{2}\ln|\mathbf{C}(\theta)|-\frac{1}{2} Z^{T} \mathbf{C}(\theta)^{-1}Z   \label{eq:margLik} 
\end{equation}
where, $ \mathbf{C}(\theta), $ is the $ \textrm{N} \times \textrm{N} $ covariance matrix with entries obtained by evaluating the chosen covariance function between all possible pairs of nominal surface points, and $ |.| $ is the matrix determinant operator. 

The method of obtaining the optimum covariance parameters by maximising the log-likelihood is robust and immune to overfitting compared to the method of obtaining the parameters by the fitting of empirical semivariogram used in Kriging \citep{MacKay1992,Stein2012interpolation}. In contrast to modelling the non-ideal behaviour through  linear or non-linear regression models \citep{Zhang2012,Schleich2014c}, which can fit limited shapes for given parameters, a given set of covariance function parameters can simulate a large number of non-ideal parts. 

The computational complexity of calculating $ \mathbf{C}(\theta)^{-1} $ scales as $\mathcal{O}(\textrm{N}^3)$, therefore, optimisation the covariance parameters for large parts where $ \textrm{N} $ is very large leads to computational issues. To overcome computational issues, in this paper, minimisation of the negative log likelihood function (equivalent to maximising log likelihood) is carried out by considering the non-ideal deviations of just the key points in Eq.~\eqref{eq:margLik}. 
Finally, the optimum values of $ \theta $ minimising Eq.~\eqref{eq:margLik}  is obtained by conjugate gradient method using the routines provided in \cite{Rasmussen2010}. This method of finding optimum values of the covariance function by minimising the negative log likelihood, eliminates the need for manual parameter guessing, and enables automatic deviation pattern modelling. For parts with more than 10000 key-points Full Independent Training Conditional (FITC) approach can be utilised and similar to the earlier case, the minimum of the negative log-likelihood function can be found by using the routines provided in \cite{Rasmussen2010}.


Additionally, when a batch of part deviation data along with key parameters which influence the part deviation such as the manufacturing process parameters or the material composition are available; trends in covariance function parameters can be identified by fitting models to capture the functional relationship between process parameters and non-ideal part behaviour. The developed functional relationship can then be used to simulate non-ideal behaviour of parts for which neither measurement nor simulation data is available for the current part. This capability of the proposed methodology to learn from historical data is demonstrated in Section~\ref{ssec:noDataCase}. 

Since the correlation lengths of the covariance function describe the behaviour of the non-ideal part along different coordinate directions when estimated using historical data they have to be transformed to match the orientation current part. This transformation is typically not necessary for automotive parts as they use the body coordinate system, and conventional designs in most cases have matching orientations. Following the estimation of optimum  covariance function parameters $ \theta $, the simulation of non-ideal parts is performed as described in Section~\ref{ssec:simulation} below.


\FloatBarrier

\subsection{Non-ideal part simulation} \label{ssec:simulation}

The estimation of optimum parameters $\theta$, as described in Section~\ref{ssec:modelling}, enables simulation of non ideal parts with spatial deviation patterns similar to the true manufactured part. The ability to simulate non-ideal parts which represent the designer's intent and conform to given GD\&T specifications is achieved by MGRF through conditional simulation. Non-ideal part simulation consists of (i)~key point selection, and (ii)~MGRF to simulate non-ideal parts. The two steps are schematically illustrated in Fig.~\ref{fig:nonIdealSim} and explained in detail in Sections~\ref{ssec:keyPointSel}~and~\ref{ssec:predVariation} below.

\begin{figure}[htb]
	\centering
	\includegraphics[width=0.4\textwidth]{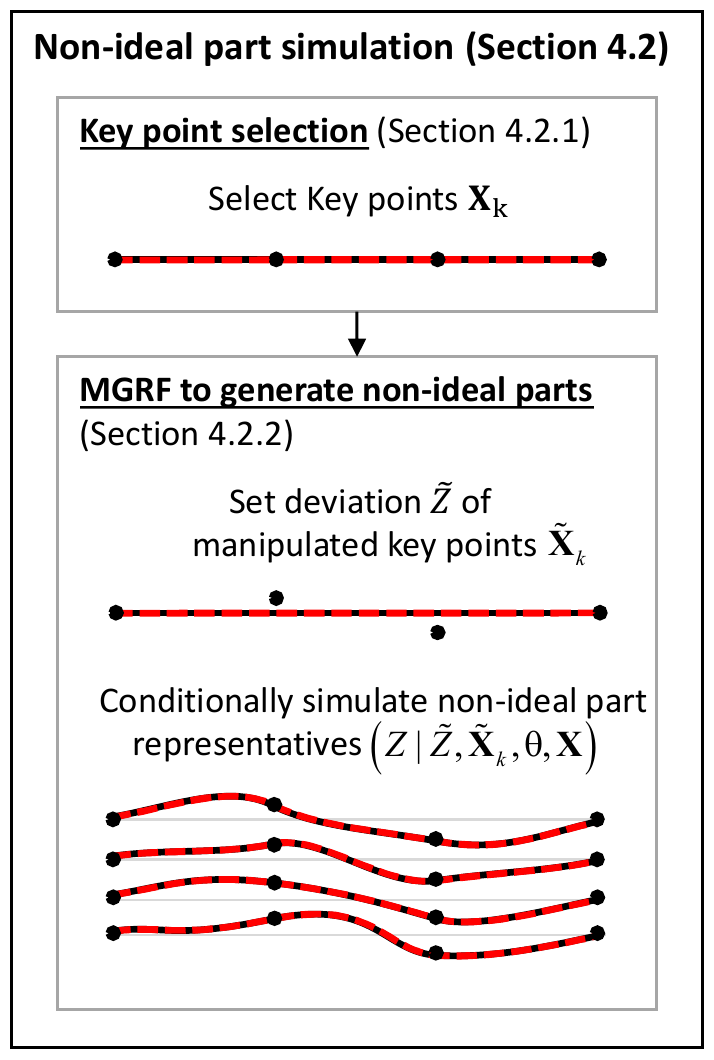} 
	\caption{Schematic representation of  non-ideal part simulation methodology}
	\label{fig:nonIdealSim}
\end{figure}

\subsubsection{Key point selection} \label{ssec:keyPointSel}

Key points $ \mathbf{X}_{k} $ are specific nodes in the mesh representation of the nominal part such that $\{\mathbf{X}_{k} \subset \mathbf{X}|{k}\in 1,2,3,...\textrm{K}\}$, where, K is the total number of key points. The generated non-ideal parts are constrained to pass through a subset of key points, $\{\tilde{\mathbf{X}}_{k} \subset \mathbf{X}_{k} |\tilde{k}\in 1,2,3,...\tilde{\textrm{K}}\}$, where, $  \tilde{\textrm{K}} $ is the total number of manipulated key points. Which are chosen to be in the region of interest, for instance on the flanges or around slots and holes. The optimal distance between key points is a function of the correlation length parameter 
\cite{Wolff2013}. A large correlation length implies a large region of influence for a given key point, therefore the key points can be spaced further apart. A large number of closely spaced key points provide the ability to precisely control the non-ideal behaviour of the part, however it entails additional computational effort. Following the selection of key points, non-ideal parts are generated by MGRF as described in Section ~\ref{ssec:predVariation} below.

\subsubsection{MGRF to generate non-ideal parts} \label{ssec:predVariation}

Simulation of non-ideal parts by MGRF consists of (i)~key points deviation setting, to reflect design intent, and (ii)~conditional simulation of non-ideal parts for a given setting of key points.

\begin{enumerate} [(i)]
	\item \textit{Key points deviation setting:} 
	
	Global and local deformation patterns are realised by manipulating the key points and setting the magnitude of their deviations. Manipulating a few key points simulates local deformation patterns, whereas, manipulating all or a large number of key points simulates global deformation patterns. The magnitude of deviations $\tilde{Z}$ and the number of manipulated key points $ \tilde{\textrm{K}} $  depend on design intent such as simulation of bending, local dent or flange geometric variation. The method of setting key point deviations is intuitive to the designer and various practical examples are illustrated in Section~\ref{ssec:dataCase}. %
	
	\item \textit{Conditional simulation of non-ideal parts:} 
	
	Conditional simulation of non-ideal parts is the two-step process of (a)~predicting the mean part deviations $(\bar{Z})$ for given setting of key points' deviations, and, (b)~generating multiple non-ideal parts conforming to given GD\&T specifications and design intent.	A detailed explanation of the steps for conditional simulation is given below as follows:	
	
	\begin{enumerate} [(a)]
		\item \textit{Prediction of mean part deviation:} 
		
		The decomposition of part form error into mean shape (systematic deviations, $ \bar{Z} $) and correlated shape error (random form error, $\xi $) as represented by Eq.~\eqref{eq:formDecomp} is the standard procedure to generate skin model shapes \citep{Schleich2014c,Yan2018}. In this paper the systematic deviations $ \bar{Z} $ as described in Section~\ref{sec:introduction} are further classified into local and global deviations that enable simulation of design intent.  		
		\begin{equation} 
		Z=\bar{Z} + \xi \label{eq:formDecomp}
		\end{equation}

		In this paper, the mean part deviations $(\bar{Z})$ for given setting of key points, is estimated by Gaussian Process Regression (GPR) \citep{Rasmussen2006gaussian}. 
		GPR is similar to the geostatistical Kriging approach \citep{ChilesJeanBook}. However, GPR considers the probabilistic interpretation of the model and that of the individual parameters of the covariance function \citep{Mackay1998introduction}, which aid in characterising the non-ideal part deviation as detailed in Section~\ref{ssec:modelling}.

		The mean non-ideal part or the expected deviation of all mesh nodes of the part $\bar{Z}$, given the deviation of key nodes, $\tilde{Z}$,  is found using Eq.~\eqref{eq:gprMean} \citep{Rasmussen2006gaussian}. 
		\begin{align}
		\bar{Z}&=\mathbb{E}(Z|\tilde{Z},\tilde{\mathbf{X}}_k, \theta, \mathbf{X}) \nonumber \\
		 &=  \mathbf{C}(\mathbf{X},\mathbf{\tilde{X}}_k) [(\mathbf{C}(\mathbf{\tilde{X}}_k,\mathbf{\tilde{X}}_k)+\sigma_{n}^{2}\mathbf{I}]^{-1} \tilde{Z}  \label{eq:gprMean}
		\end{align}
		where,  $\mathbf{X}$ is the $\textrm{N}\times3$ matrix of nominal coordinates of all mesh nodes; $\mathbf{\tilde{X}}_k$, is the $\tilde{\textrm{K}}\times3$ matrix of nominal co-ordinates of manipulated key nodes; $\mathbf{C}(\mathbf{X},\mathbf{\tilde{X}}_k)$, is the $\textrm{N}\times\tilde{\textrm{K}}$ covariance matrix with the chosen covariance function evaluated between all pair of points in $ \mathbf{X} $ and  $\mathbf{\tilde{X}}_k $, similarly $\mathbf{C}(\mathbf{\tilde{X}}_k,\mathbf{\tilde{X}}_k)  $, is the $\tilde{\textrm{K}}\times\tilde{\textrm{K}}$ covariance matrix; $\mathbf{I}$, is an identity matrix with size equal to the number of manipulated key points; $ \sigma_{n}^{2} $, traditionally is the nugget effect or measurement error variance \citep{Rasmussen2006gaussian,ChilesJeanBook}, which in this paper is set to zero so that the generated non-ideal parts pass through the key points, $\mathbf{\tilde{X}}_k$. The prediction of the mean part using GPR has the following advantages: (i)~it does not require guessing the degree and order of the regression, and, (ii)~it does not limit the non-ideal behaviour of parts to second order shapes.
		
		\item \textit{Simulation of multiple non-ideal parts:}
		

		In this paper, multiple non-ideal parts are simulated by conditional simulations \citep{ChilesJeanBook} with modifications to (i)~simulate parts conforming to GD\&T form tolerance specifications, and (ii)~overcome computational limitations that entail non-ideal part simulation of large parts. The methodology utilises unconditional simulations to generate  non-ideal parts conforming to given GD\&T specification and design intent. Therefore, firstly, the methodology to generate an unconditional simulation of spatially correlated random variables is detailed. Following which, the methodology to utilise the generated unconditional simulations for the simulation of non-ideal parts conforming to given GD\&T specification and design intent is detailed.  
		
		The unconditional simulation of an  N$\times$1 vector $\xi_u$, with a spatial deviation pattern similar to the true manufactured part can be generated by sampling from the N$\times$N covariance matrix, $ \mathbf{C}(\mathbf{X},\mathbf{X}) $, obtained using optimum parameters $ \theta $ estimated in Section~\ref{ssec:modelling}. The sampling can be performed by the multiplication of a standard Gaussian random vector (U) with (i)~Cholesky factor of the covariance matrix as in Eq.~\eqref{eq:covFactor1}, or, (ii)~eigen-decomposition of the covariance matrix  as in Eq.~\eqref{eq:covFactor2} \citep{Davis1987a}. 
		\begin{subequations} \label{eq:covFactor}
			\begin{align} 
			\xi_u &= \mathbf{L} \textrm{U} \label{eq:covFactor1} \\
			&= \mathbf{\Phi} \mathbf{\Lambda}^{\frac{1}{2}} \textrm{U} \label{eq:covFactor2}
			\end{align}		
		\end{subequations}
		where, $\mathbf{L}$ is the N$\times$N lower triangular Cholesky  factor of~$ \mathbf{C}(\mathbf{X},\mathbf{X}) $, $\mathbf{\Phi}$ is the N$\times$N eigen basis matrix of~$\mathbf{C}(\mathbf{X},\mathbf{X}) $, and, $\mathbf{\Lambda}$ is the diagonal eigenvalue matrix of~$\mathbf{C}(\mathbf{X},\mathbf{X}) $. Multiple instances of  $\xi_u$ can be obtained by generating independent instances of $\textrm{U}$ and substituting it in Eq.~\eqref{eq:covFactor}.

		Cholesky decomposition or eigen-decomposition  of the N$\times$N covariance matrix, $\mathbf{C}(\mathbf{X},\mathbf{X})$, is computationally challenging, when N, the number of mesh nodes becomes greater than 10,000. To overcome this limitation, the eigen-decomposition is performed on the K$\times$K covariance matrix $\mathbf{C}_\textrm{K}$, obtained by limiting $\mathbf{X}$ in $\mathbf{C}(\mathbf{X},\mathbf{X})$ to the key nodes $\mathbf{X}_k $, to obtain the key point eigen basis matrix,~$\mathbf{\Phi}_\textrm{K}$. The eigen basis matrix $\mathbf{\Phi}$ for all mesh nodes of the is estimated by interpolating the values at key nodes \citep{Wolff2013}. Is this paper the interpolation of eigen basis $\mathbf{\Phi}$ from $\mathbf{\Phi}_\textrm{K}$, is performed by solving the Poisson  problem Eq.~\eqref{eq:poission} using FEM formulation utilising routines provided in \citep{Franciosa2018}, however, this can be performed by utilising any interpolation method. 
		\begin{equation} \label{eq:poission}
		\left(\frac{\partial^2}{\partial x^2}+ \frac{\partial^2 }{\partial y^2}+\frac{\partial^2}{\partial z^2}\right)\mathbf{\Phi}= \mathbf{\Phi}_\textrm{K}
		\end{equation}

		Typically the first $ \textrm{R} $ columns of $\mathbf{\Phi}_\textrm{K}$, $ \textrm{R} \ll \textrm{K} $, account for most of the variance in data and $ \xi_u $ can be generated using Eq.~\eqref{eq:intepCor}. 
		\begin{equation} 
		\xi_u =\mathbf{\Phi}_\textrm{R} \mathbf{\Lambda}^{\frac{1}{2}}_\textrm{R} \textrm{U}_\textrm{R} \label{eq:intepCor}
		\end{equation}
		where,  $\mathbf{\Phi}_\textrm{R} $ is the N$\times$R interpolated eigen basis matrix, $ \mathbf{\Lambda}_\textrm{R} $ is the R$\times$R diagonal matrix with elements equal to the first R eigenvalues of $\mathbf{C}_\textrm{K}$, and $ \textrm{U}_\textrm{R} $ is the R$ \times $1 vector of independent standard Gaussian random variables. Thus, enabling the unconditional simulation of spatially correlated patterns for large non-ideal parts.

		The Conditional simulation of a non-ideal part conforming to GD\&T and design intent is obtained by first estimating the variance or the scaling factor parameter $\sigma_{T} $, of the covariance function that corresponds to a given form tolerance specification. As described in Section \ref{sec:introduction}, in this paper, we demonstrate the simulation of non-ideal parts conforming to the specified form tolerance requirements for the profile of a surface \cite{ISO-TC2013}, specifically the free state tolerance specification as described in \cite{ISO105792013}. 
		
		Considering a statistical tolerance CI of $ p \% $, Lower Specification Limit (LSL) and Upper Specification Limit (USL) of $ t/2 $, i.e., $ |\textrm{LSL}|=|\textrm{USL}|=t/2 $, we estimate $ \sigma_{T}  $ using Eq.~\ref{eq:sigmaT}.		
		\begin{equation}
		\sigma_{T} = \frac{|\textrm{USL}|}{S_z} \label{eq:sigmaT}
		\end{equation}
		where, $ S_z $ is the Z-score for a standard normal distribution corresponding to cumulative probability $P\leq (1+p)/2 $. Following the estimation of $ \sigma_{T} $, steps described in detail in lines 3-10 of Algorithm~\ref{alg:formTolTrans} lead to non-ideal parts that conform to both GD\&T specifications and design intent.

	\end{enumerate}
	
\end{enumerate}

In contrast to existing methodologies which add random form deviation, the form deviation simulated through the proposed MGRF methodology has the following advantages: (i)~it is designer centric as it enables intuitive and precise control of non-ideal part deviations, and, (ii)~it simulates technological patterns similar to the source deviation, a capability lacking existing designer centric methodologies. 
A demonstration of the capabilities  of the developed MGRF methodology is presented in Section~\ref{sec:Case1} below.
\begin{algorithm}
	\KwData{Max form error magnitude ($|\textrm{USL}|$), statistical tolerance conformance probability ($ p $), optimised covariance function parameters $(\theta)$, key point deviations$ (\tilde{Z}) $ }	
	
	\KwResult{Morphed non-ideal parts conforming to given GD\&T specifications and design intent.}
	
	Calculate Z-score $ S_z $ for a standard normal distribution corresponding to cumulative probability $P\leq (1+p)/2 $\;
	
	Calculate the standard deviation $ \sigma_{T} $ of a zero mean normal distribution for which the probability of a random variable having value  $ \leq |\textrm{USL}| $ is $ p $, by using Eq.~\eqref{eq:sigmaT}\;
	
	Generate mean surface $ \bar{Z} $ by setting key point deviations to $ \tilde{Z} $, using GPR Eq.~\eqref{eq:gprMean} \;
	
	\For{each non-ideal part representative to be simulated}{

		Generate scaled unconditional simulation $ \xi_u $, by sampling from covariance matrix obtained by using optimised correlation length parameters from   $\theta$ and setting $ \sigma_{f} = \sigma_{T} $\;		
		
		Find the deviation value at key points in $ \xi_u $\;
		
		Generate a surface $ \bar{Z}_k $ using GPR Eq.~\eqref{eq:gprMean}, with deviations at key points equal to the value found in line 6\;
		
		Find the difference in deviation at each node between the un-conditioned surface generated in line 5 and mean surface generated in line 7, i.e. $ \bar{Z}_k-\xi_u $\;

		Add the difference obtained in line 8 at each node to the corresponding node of the mean surface $ \bar{Z} $ generated in line 3, to obtain a non-ideal part conforming to given form tolerance specification and design intent, i.e. $ Z=\bar{Z}+\bar{Z}_k-\xi_u $\;
	}
	
	\caption{Conditional simulation of non-ideal parts conforming to GD\&T specifications and design intent}
	\label{alg:formTolTrans}
\end{algorithm}

%% file: caseStudy.tex
\section{Applications in automotive assembly process} \label{sec:Case1}

The MGRF methodology developed in this paper is demonstrated using sport utility vehicle door inner sheet-metal parts. The door inner is a vital part of the door subassembly which consists of window channel, halo, hinge, latch reinforcement, and seat belt reinforcement. As described in Section~\ref{sec:introduction}, non-ideal behaviour of parts play a key role in determining quality of joining/fastening 
with other parts belonging to the subassembly. 
Therefore, its effect on assembly quality has to be quantified as early as possible through simulation studies.  

In this section, defect fidelity, designer centricity, and ability  to support tolerance analysis and synthesis are illustrated for two specific scenarios of data availability common during early design phase, demonstrating the MGRF methodology's data versatility. Firstly, we demonstrate the application of MGRF methodology to the case where measurement or simulation data for the current part is available, representing the engineering and assembly stage of the NPI process of an assembly system development. Secondly, we consider the case where data for the current part is unavailable but historical data for a similar part is available, representing the engineering phase of the same NPI process. 
The two scenarios are explained in detail in Sections~\ref{ssec:dataCase}~and~\ref{ssec:noDataCase}.

\subsection{Case 1: Measurement or simulation data for the current part are available} \label{ssec:dataCase}

\subsubsection{Non-ideal part modelling} \label{sssec:modWithData}

\begin{figure}[htb] 
		\centering	
		\includegraphics[width=\columnwidth]{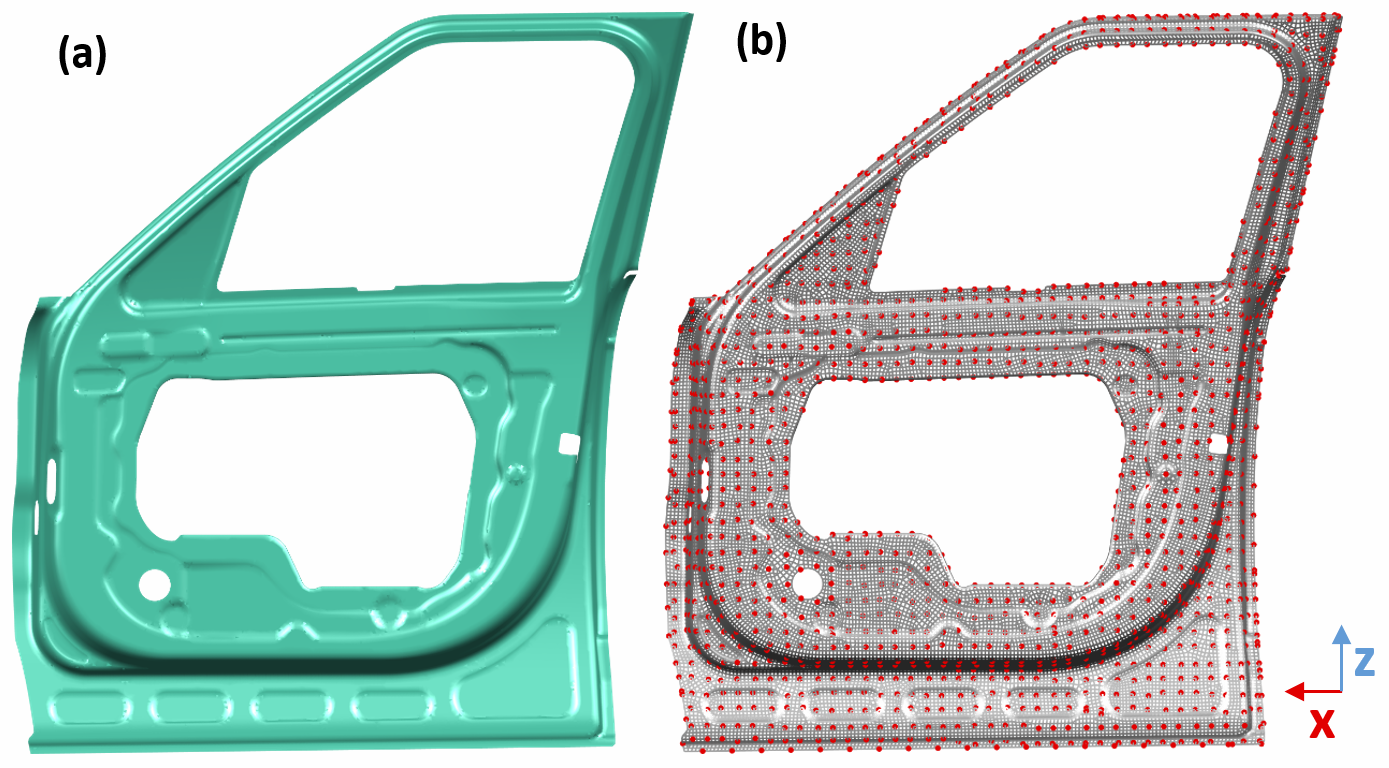}
		\caption{(a)~CAD geometry, (b)~mesh representation with key nodes}
		\label{fig:compoonentMeshes}
	\end{figure}

	\textit{Input and pre-processing:}
	The inputs, in this case, are the CAD model of the ideal geometry of the part and measurement data. The CAD model of the part is converted to its mesh representation using Altair Hypermesh commercial software; the CAD along with its discrete meshed representation is illustrated in Fig.~\ref{fig:compoonentMeshes}. The mesh representation of the door inner consists of 75,000 mesh nodes and demonstrates the scalability of the proposed methodology to large parts.	
	
\begin{figure}[htb] 
		\centering	
		\includegraphics[width=\columnwidth]{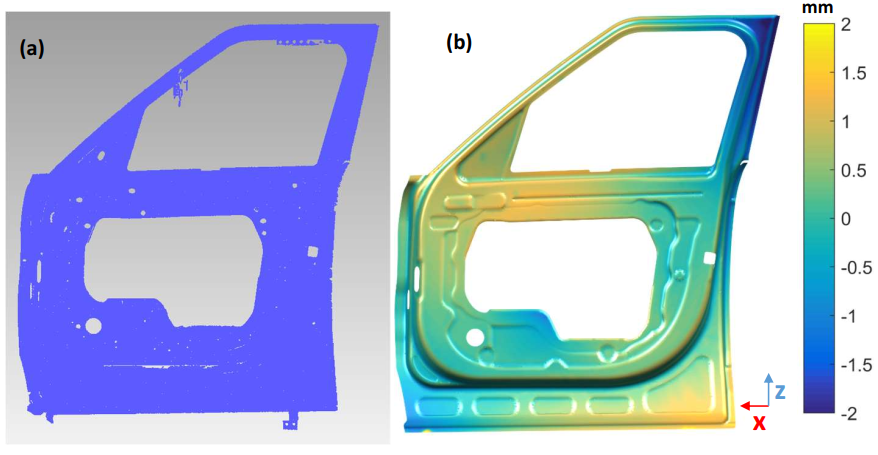}  
		\caption{(a)~measured Cloud-of-Points, (b)~estimated part surface normal deviation}
		\label{fig:componentVariations}
		\end{figure}

	The true manufactured part's measurement data in the form of CoP is obtained from an optical 3D-surface scanner and is illustrated in Fig.~\ref{fig:componentVariations}a. Utilising this CoP data the deviation of each mesh node ($Z$) in the surface normal direction is calculated by routines available in \citep{Franciosa2018}. The colour map of the obtained surface normal deviation is shown in Fig.~\ref{fig:componentVariations}b.
	
	\textit{Characterisation of non-ideal part deviation:}	
	Since no small scale variation or periodicity was observed in the measurement data, a single  Mat\'ern covariance function was used to characterise the deviation field. The Mat\'ern covariance function is utilised due to its computational advantages compared to the squared exponential function \cite{Stein2012interpolation}. The optimum parameters $ \theta $ were found by maximising the likelihood (Eq.~\eqref{eq:margLik}) and the optimum values of the correlation length in X, Y and Z coordinate directions were estimated to be 168.73 mm, 7.74 mm, and 93.70 mm, respectively. The obtained optimum correlation lengths reflect the true deviation pattern seen in Fig.~\ref{fig:componentVariations}b, where a given non-ideal deviation in X coordinate direction propagates longer compared to the deviations in Y and Z coordinate directions. Following the estimation of optimum covariance function parameters, the simulation of non-ideal parts is carried out as described in Section~\ref{sssec:simWithData} below.


\subsubsection{Non-ideal part simulation} \label{sssec:simWithData}

	\textit{Key points selection:} 	
	The key points in this paper are chosen uniformly from all the mesh nodes of the part by finding the points of intersection of the part and its voxalised bounding box with cubic voxels of size 20 mm. The distance of 20 mm between two points is smaller than the optimum correlation lengths in X and Z coordinate directions enabling increased control on generated non-ideal part instances. The selected key mesh nodes for the part are coloured red and superimposed on the mesh representation in Fig.~\ref{fig:compoonentMeshes}b. 
	
	\textit{MGRF to generate non-ideal parts:}	
	This section illustrates the designer centricity of MGRF methodology through the simulation of non-ideal parts for the following `what if?' scenarios: (i)~global deformation patterns, (ii)~local deformation patterns, and (iii)~parts that conform to GD\&T form tolerance specification for the profile of a surface. The aforelisted scenarios can be simulated by simple manipulation of key-points which adds to the designer centricity of the methodology.
	
\begin{figure}[htb] 
		\centering	
		\includegraphics[width=\columnwidth]{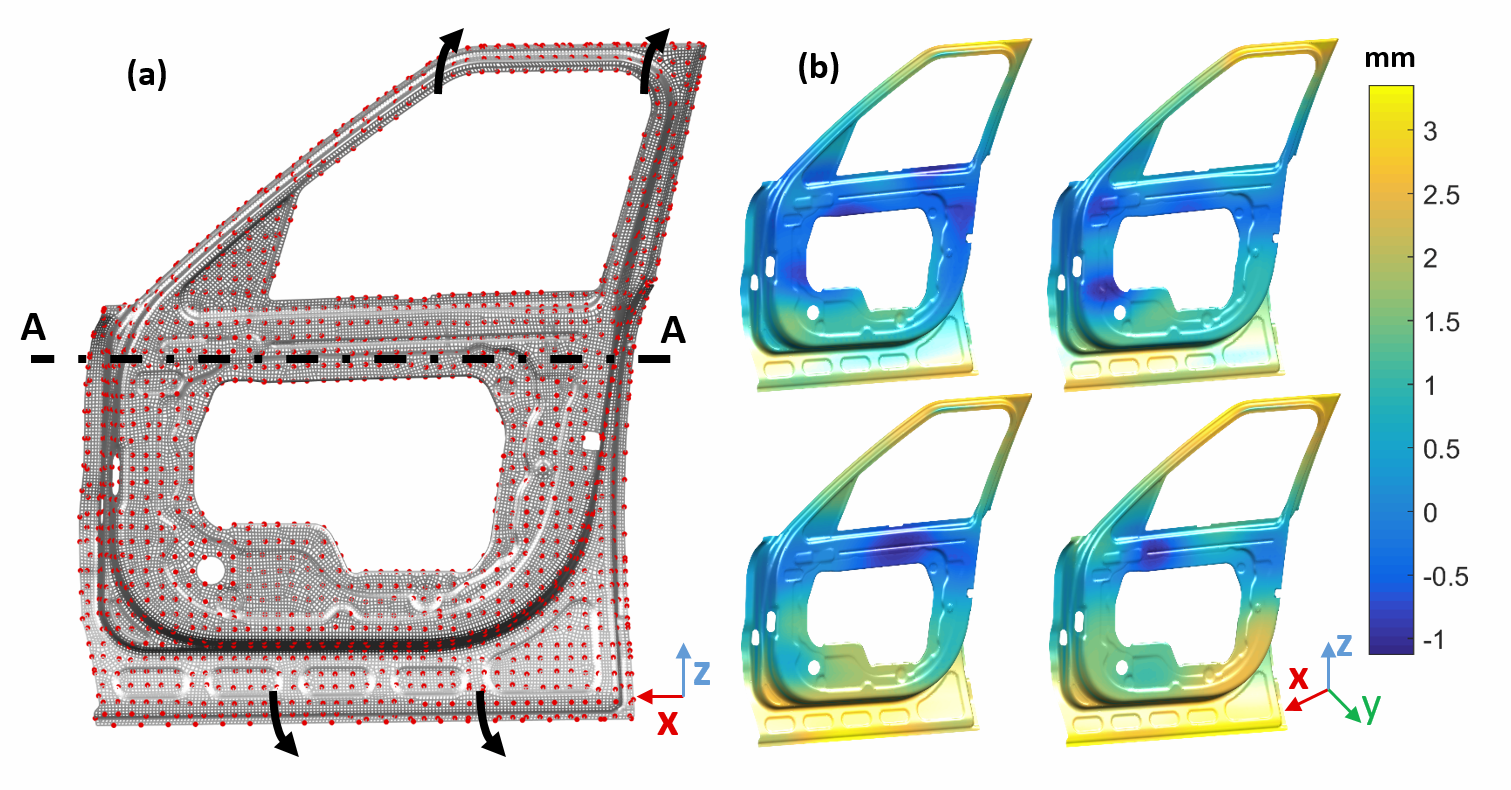}
		\caption[Illustration of gobal non-ideal part deformation]{Illustration of global non-ideal part deformation (a)~key points' setting, (b)~simulated part instances }
		\label{fig:golbalDef}
	\end{figure}

	Firstly, global deformation patterns of the non-ideal part are simulated by manipulating a large number of key points according to the design intent. In this section, a bending of the part about axis AA is simulated (Fig.~\ref{fig:golbalDef}a), giving rise to a maximum deviation of 3mm. The mean part shape simulating bending is found through GPR as described in Section~\ref{ssec:predVariation}-(ii)-(a) by setting all key points to proportionately deviate about axis AA, with the farthest key point from axis being deviated by 3mm. A form error with a magnitude of  $ \pm1 $ mm and a statistical tolerance CI of 95\% obtained by conditional simulation as described in Section~\ref{ssec:predVariation}-(ii)-(b) is added to the mean bending shape. Four generated non-ideal part instances are shown in Fig.~\ref{fig:golbalDef}b.
	
\begin{figure}[htb] 
		\centering	
		\includegraphics[width=\columnwidth]{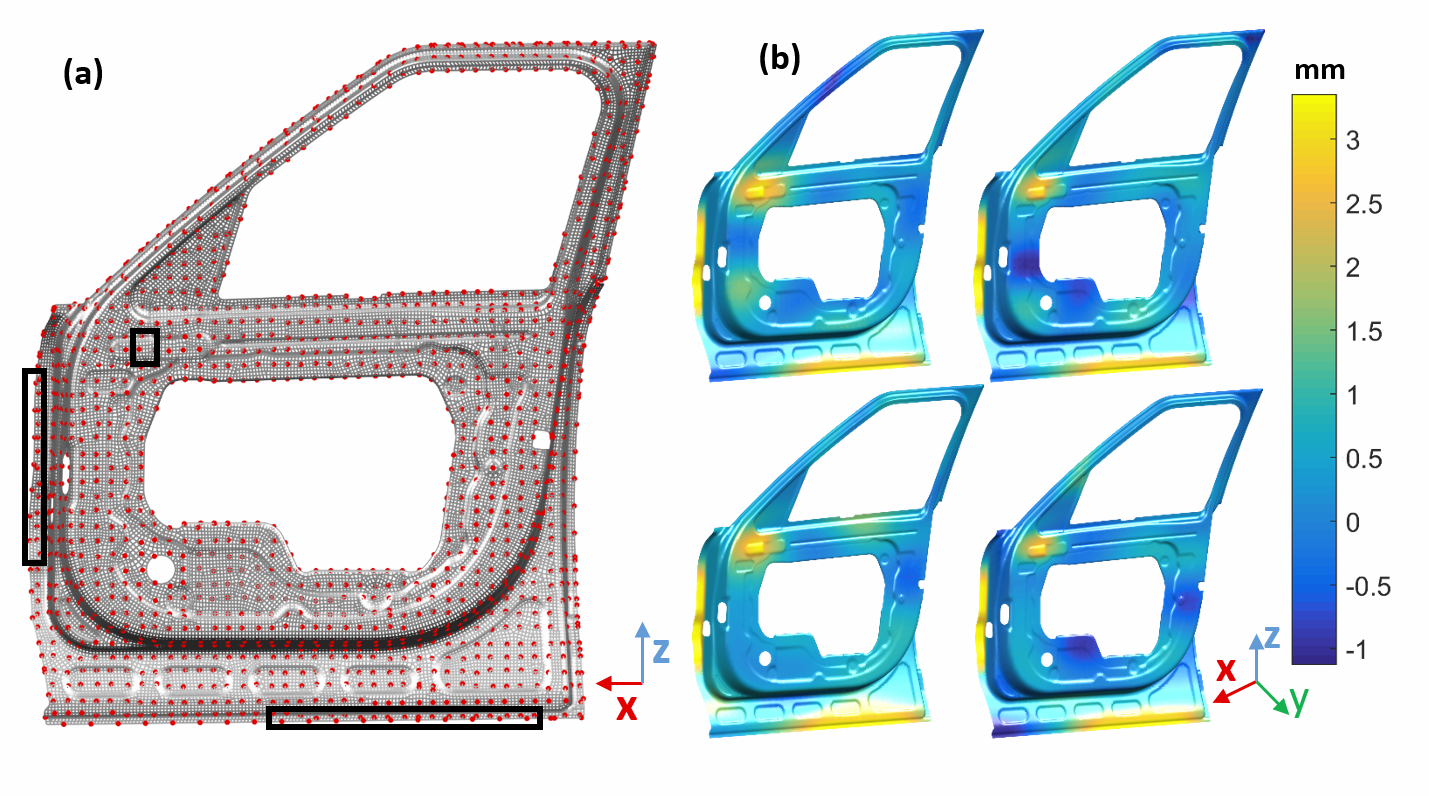}
		\caption[Illustration of local non-ideal part deformation]{Illustration of local non-ideal part deformation (a)~key points' setting, (b)~simulated part instances}
		\label{fig:localDef}
	\end{figure}

	Secondly, local deviations of the non-ideal part are simulated by manipulating few key nodes corresponding to the type of deviation to be simulated. A local deviation can be a dent or a localised deformation of the flange where the joining of two parts take place, these variations are of practical importance and need to be simulated. Here both a dent and local flange variation of 3mm from the nominal are simulated by moving the key nodes within the black rectangles in Fig.~\ref{fig:localDef}a by 3mm. Following the setting of key point deviations, the mean part shape is found through GPR to which a form error for the profile of a surface with a magnitude of $ \pm1 $ mm and a statistical tolerance CI of 95\% obtained by conditional simulation is added. Various non-ideal part representatives simulated under the given local deformation constraints are shown in Fig.~\ref{fig:localDef}b. The extent to which the deviation of a key point affects the neighbouring points in a given direction is determined by the corresponding correlation length. The use of optimised correlation lengths estimated in Section~\ref{sssec:modWithData} enables automatic emulation of spatial pattern in measurement data. 
	
\begin{figure*}[htb] 
		\centering	
		\includegraphics[width=0.8\textwidth]{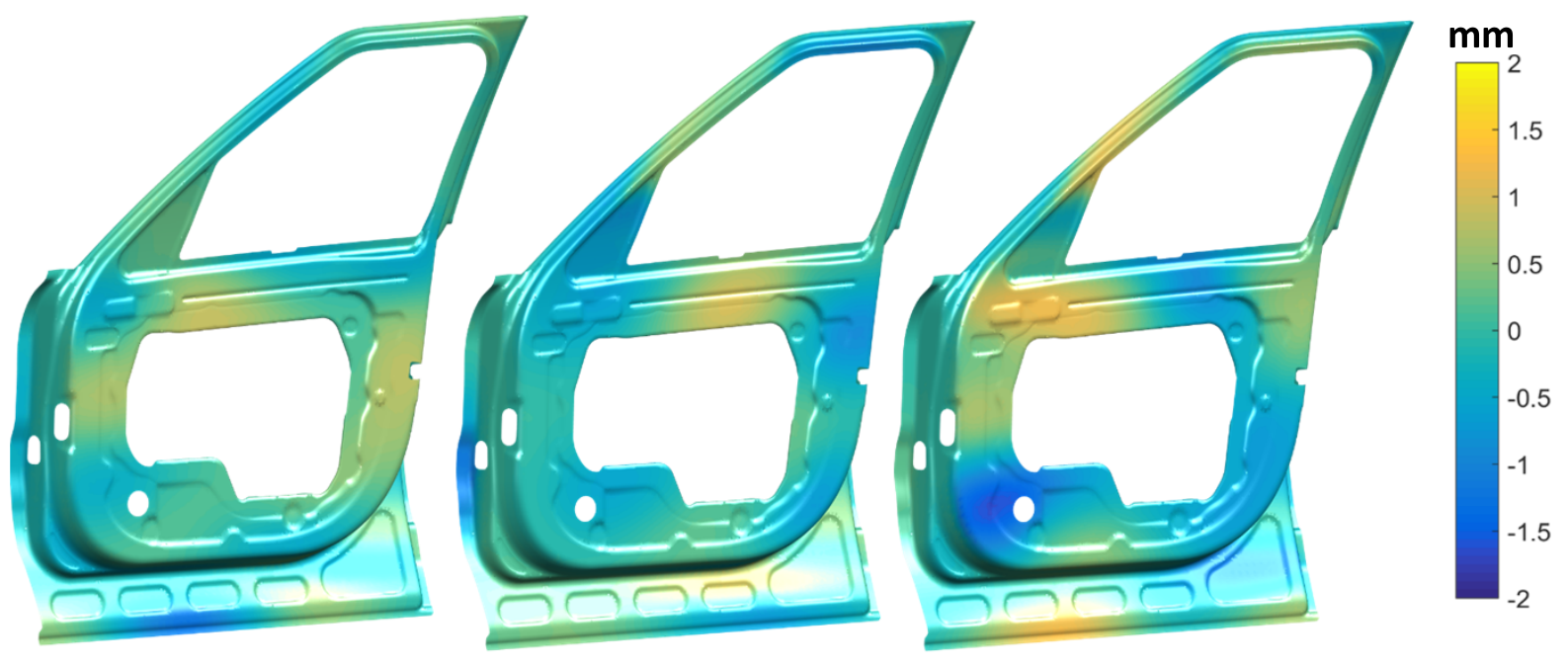} 
		\caption[Illustration of part form error simulation conforming to GD\&T]{Illustration of part form error simulation conforming to GD\&T form tolerance specification for the profile of a surface}
		\label{fig:formError}
	\end{figure*}

	Finally, non-ideal part representatives conforming to GD\&T form tolerance specification for the profile of a surface, with no global or local deformations are simulated by setting all key point deviations to zero and adding the generated scaled correlated spatial patterns obtained as described in Section~\ref{ssec:predVariation}-(ii-b) to the ideal surface. Figure.~\ref{fig:formError} illustrates three such simulated part form variations of $ \pm 2 $mm with statistical tolerance CI of 95\%.	
	

\subsection{Case 2: Historical data of similar part available} \label{ssec:noDataCase}

\subsubsection{Non-ideal part modelling} \label{sssec:modWithoutData}
\begin{figure}[htb] 
		\centering	
		\includegraphics[width=\columnwidth]{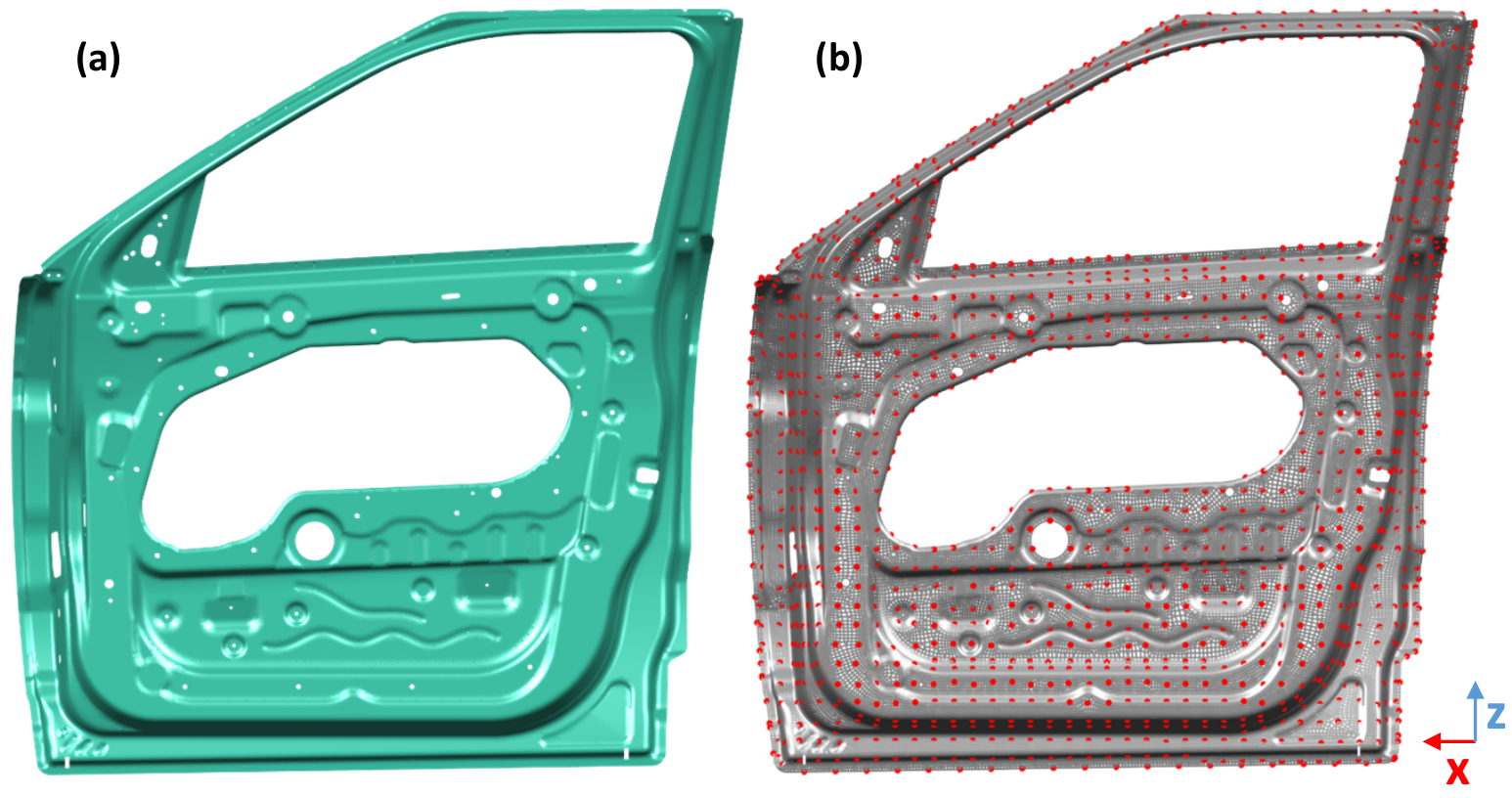} 
		\caption{(a)~CAD geometry, (b)~mesh representation with key nodes}
		\label{fig:compoonentMesheSt}
	\end{figure}

	\textit{Input and pre-processing:} 
	The inputs, in this case are (i)~the CAD model of the ideal geometry of the part, and (ii)~CAD model and historical data of a similar part. We utilise the door inner part described in Fig.~\ref{fig:compoonentMeshes} as the historically similar part with data. The pre-processing is performed as described in Section~\ref{sssec:modWithData}, the CAD geometry and mesh representation of the new part for which no real manufactured part measurement data is available is illustrated in Fig.~\ref{fig:compoonentMesheSt}.
	
	\textit{Characterisation of non-ideal part deviation:} 
	The optimum correlation function parameters obtained from historical data can be utilised for non-ideal part simulation of the part for which no data is available if the parts (i)~are geometrically similar, (ii)~have similar material composition, and (iii)~are fabricated using the same manufacturing process. In addition to the above constraints, the parts should also have the same orientation, enabling the transfer of correlation length parameters. In this case study, the transformation was not necessary as the parts utilise the same coordinate system and have the same orientations, i.e., both parts are right side door inners of similar automobile models. To demonstrate the learning from historically similar part, the optimum parameters estimated in Section~\ref{sssec:modWithData} are utilised to simulate non-ideal parts for the new part illustrated in Fig.~\ref{fig:compoonentMesheSt}.
	
\begin{figure}[htb]
		\centering
		\includegraphics[width=\columnwidth]{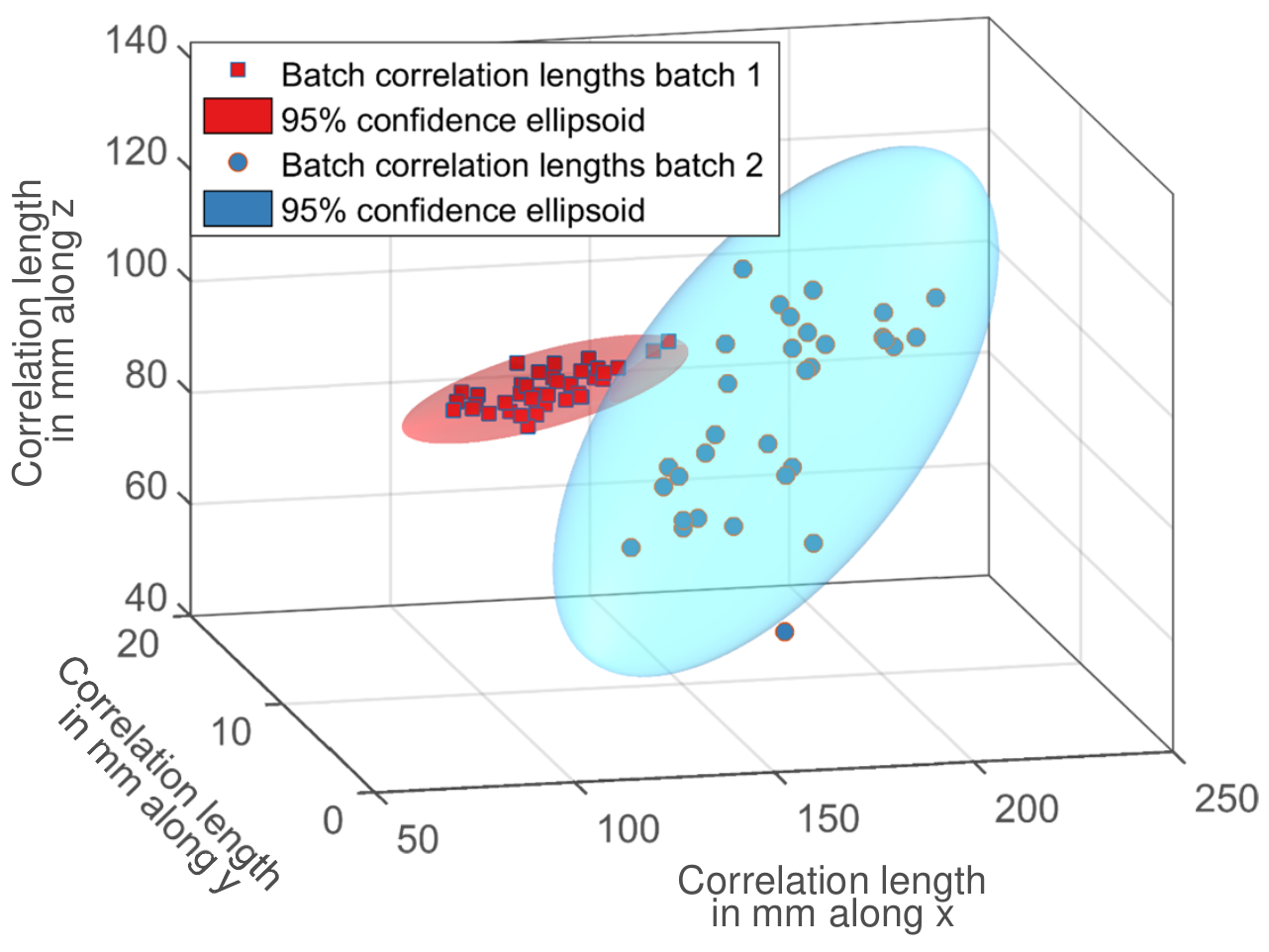} 
		\caption{Covariance function parameters characterisation}
		\label{fig:batchParam}
	\end{figure}

	Additionally, as described in Section~\ref{ssec:optimiseParam}, a functional relationship between process parameters and the optimum covariance function parameters can be modelled. The model can then be utilised to obtain optimum parameters in the future. A simple illustration of this characterisation is shown in Fig.~\ref{fig:batchParam}, where 3-Dimensional Gaussian distributions are fit to the optimum correlation lengths of two batches of manufactured parts with different materials. The correlation lengths of parts from two batches form separate clusters differentiating them. A two tailed t-test with the null hypothesis that the correlation lengths of the two batches belong to a Gaussian distribution with equal means was rejected with all p-values $ < 0.025 $. However, the same test performed on two groups obtained by splitting the correlation lengths of each batch (i.e. individual batches were divided into two equal groups and separate t-tests were performed o each batch) failed to reject the null hypothesis of equal means with all p-values $ > 0.22 $, i.e. proved that each batch belonged to the same Gaussian distribution with equal means. Thus demonstrating that correlation lengths can be used effectively as a  means to model non-ideal part characteristics.

	Finally, when no historical data are available the optimum covariance function parameters can be manually set as they are easily interpreted as the distance until which the non-ideal deviation at point on the surface can influence its surrounding along that direction. This capability adds to the designer centricity of the MGRF methodology.



\subsubsection{Non-ideal part simulation} \label{sssec:simWithoutData}

\begin{figure*}[htb] 
		\centering	
		\includegraphics[width=0.8\textwidth]{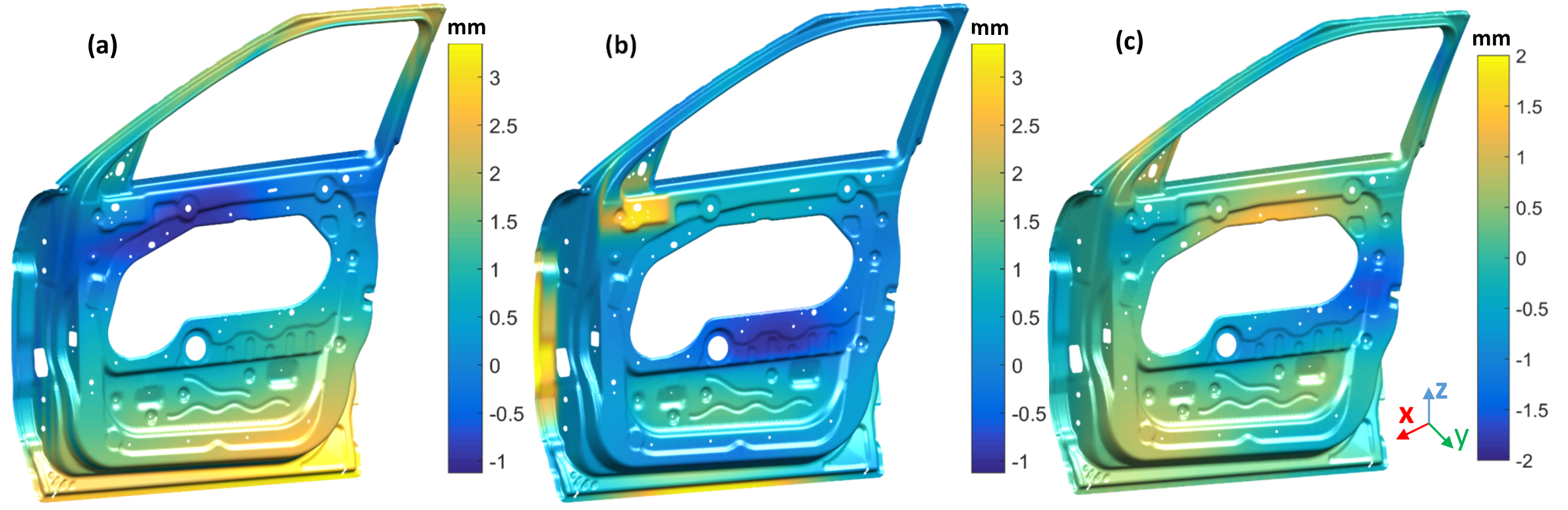} 
		\caption{Illustration of non-ideal part simulation with historical data (a)~global part deformation, (b)~local part deformation, (c)~conformance to form tolerance specification for profile of a surface}
		\label{fig:partIndepen}
	\end{figure*}

The Key points selection and MGRF to generate non-ideal parts are performed as described in Section~\ref{sssec:simWithData} and the resulting non-ideal parts simulated using optimum covariance function parameters of the historical part are illustrated in Fig.~\ref{fig:partIndepen}. Thus demonstrating that the spatial deviation pattern from historical data of a similar part can be transferred to the part with no data, a capability that is essential for non-ideal part simulation during early design stage.

%% file: conclusion.tex
\section{Conclusions and future work} \label{sec:conclusion}
	
A novel MGRF methodology for OSEMS was developed in this paper. Overcoming the limitations of State-of-Art methodologies the MGRF methodology has the following contributions that enable it to effectively model and simulate object shape error during the early design stage: (i) it has high defect fidelity so that it is capable of simulating local and global deformations, and technological patterns; (ii) it has high data versatility enabling it to effectively simulate non-ideal parts at all levels of data availability; (iii)  it is highly designer centric, capable of performing `what if?' analysis, and has model parameters that are physically meaningful; and (iv) it supports simulation of statistical form tolerance for profile of a surface without additional modelling effort. All the aforementioned capabilities demonstrated through industrial case studies in Section~\ref{sec:Case1} clearly illustrate the contributions of the MGRF methodology.

Though the MGRF methodology has many advantages it entails a few limitations, which are as follows: (i)~non-ideal part deviations are assumed to be in the surface normal direction, this assumption could lead to inaccurate modelling of non-ideal behaviour in the regions of high curvature; (ii)~dependence on designer for key point deviations, this could lead to generation of unrealistic non-ideal part representatives; and (iii)~inability to simulate unsymmetrical tolerance bounds and variations that are non-Gaussian. Future scope of research includes developing a clear methodology to identify the scope of parts' geometric similarity to utilise historical data. Though the ability to effectively utilise historical data has been demonstrated in this paper, a few research questions are yet to be addressed. For instance, a definition of similarity of parts in terms of topology of the parts or the manufacturing process used to fabricate the parts, has to be provided.